\documentclass[%
prd,nofootinbib,showpac,
superscriptaddress,
preprint
]{revtex4-1}
\usepackage{adjustbox}
\usepackage{graphicx}
\usepackage{dcolumn}
\usepackage{bm}

\usepackage[utf8x]{inputenc} 
\usepackage{amsmath,amssymb} 
\usepackage{adjustbox}
\usepackage{color}

\usepackage[colorlinks,citecolor=blue]{hyperref}

\usepackage{float}
\usepackage{rotating}
\usepackage[font=small,labelfont=bf]{caption}
\usepackage{upgreek}
\usepackage{subcaption}

\newcommand{\nue}{$\nu_{e}$}
\newcommand{\numu}{$\nu_{\mu}$}
\newcommand{\nueb}{$\bar{\nu}_{e}$}
\newcommand{\thetamu}{$\theta_{23}$}
\newcommand{\numub}{$\overline{\nu}_{\mu}$}

\newcommand{\dmatm}{$\Delta m^2_{31}$}
\newcommand{\absdmatm}{$|\Delta m^2_{31}|$}

\newcommand{\dcp}{$\delta_{\text{CP}}$}

\newcommand{\nova}{NO$\nu$A}
\newcommand{\numode}{$\nu$-mode}
\newcommand{\antinumode}{$\bar{\nu}$-mode}
\newcommand{\conlev}{C.L.}

\begin{document}

\preprint{arXiv:2009.08585 [hep-ph]}

\title{Physics potential of the combined sensitivity of T2K-II, NO$\nu$A extension, and JUNO}

	\author{S. Cao}
	\email{cvson@ifirse.icise.vn}
	\affiliation{\textit{Institute for Interdisciplinary Research in Science and Education,\\ \it{ICISE, Quy Nhon, Vietnam.}}}
	\affiliation{\textit{High Energy Accelerator Research Organization (KEK), Tsukuba, Ibaraki, Japan.}}
	\author{A. Nath}
	\email{ankur04@tezu.ernet.in}
	\affiliation{\textit{Department of Physics, Tezpur University, Assam, India.}}
	\author{T. V. Ngoc}
    \email{tranngocapc06@ifirse.icise.vn}
    \affiliation{\textit{Institute for Interdisciplinary Research in Science and Education,\\ \it{ICISE, Quy Nhon, Vietnam.}}}
	\affiliation{\textit{Graduate University of Science and Technology, Vietnam Academy of Science and Technology, Hanoi, Viet Nam.}}
	\author{Ng. K. Francis}
	\affiliation{\textit{Department of Physics, Tezpur University, Assam, India.}}
	\author{N. T. Hong Van}
	\affiliation{\textit{Institute of Physics, Vietnam Academy of Science and Technology, Hanoi, Vietnam.}}
	\author{P. T. Quyen}
    \affiliation{\textit{Institute for Interdisciplinary Research in Science and Education,\\ \it{ICISE, Quy Nhon, Vietnam.}}}
	\affiliation{\textit{Graduate University of Science and Technology, Vietnam Academy of Science and Technology, Hanoi, Viet Nam.}}

\date{\today}

\begin{abstract}
Leptonic \textit{CP} violation search, neutrino mass hierarchy determination, and the precision measurement of oscillation parameters for a unitary test of the leptonic mixing matrix are among the major targets of the ongoing and future neutrino oscillation experiments. The work explores the physics reach for these targets by around 2027, when the third generation of the neutrino experiments starts operation, with a combined sensitivity of three experiments: T2K-II, \nova\ extension, and JUNO. It is shown that a joint analysis of these three experiments can conclusively determine the neutrino mass hierarchy. Also, at certain values of \emph{true} \dcp, it provides closely around a $5\sigma$ confidence level (C.L.) to exclude \textit{CP}-conserving values and more than a $50\%$ fractional region of \emph{true} \dcp\ values can be explored with a statistic significance of at least a $3\sigma$ C.L. Besides, the joint analysis can provide unprecedented precision measurements of the atmospheric neutrino oscillation parameters and a great offer to solve the \thetamu\ octant degeneracy in the case of nonmaximal mixing.  
\end{abstract}

\maketitle


\section{\label{intro}Introduction}
Neutrino oscillation, discovered by the Super-Kamiokande (SK) experiment~\cite{fukuda1998evidence} and the Sudbury Neutrino Observatory~\cite{Ahmad:2001an,ahmad2002direct}, establishes palpable evidence beyond the description of the Standard Model of elementary particles: neutrinos have masses and the leptons mix. This phenomenon is described by a $3\times3$ unitary matrix, widely known as the Pontecorvo-Maki-Nakagawa-Sakata (PMNS)~\cite{maki1962remarks,pontecorvo1968neutrino} matrix, which connects three neutrino flavor eigenstates $(\nu_e,\nu_{\mu},\nu_{\tau})$ with three neutrino mass eigenstates $(\nu_1,\nu_2,\nu_3)$ given by a corresponding mass spectrum $(m_1,m_2,m_3)$. The matrix is commonly parametrized by three leptonic mixing angles ($\theta_{12},\theta_{13},\theta_{23}$), one \textit{CP}-violating phase $(\delta_{\text{CP}})$, and two Majorana phases ($\rho_{1}$, $\rho_{2}$), and can be written as
\begin{equation*}
\resizebox{0.5\textwidth}{!}{$
    U_{\text{PMNS}}=
    \begin{pmatrix}
    c_{12}c_{13} & s_{12}c_{13} & s_{13}e^{-i\delta_{\text{CP}}}\\
    -s_{12}c_{23}-c_{12}s_{13}s_{23}e^{i\delta_{\text{CP}}} & c_{12}c_{23}-s_{12}s_{13}s_{23}e^{i\delta_{\text{CP}}} & c_{13}s_{23}\\
    s_{12}s_{23}-c_{12}s_{13}c_{23}e^{i\delta_{\text{CP}}} & -c_{12}s_{23}-s_{12}s_{13}c_{23}e^{i\delta_{\text{CP}}} & c_{13}c_{23}
    \end{pmatrix}P_{m}$,}
    \label{eqPM}
\end{equation*}
where $c_{ij}=\cos{\theta_{ij}}$, $s_{ij}=\sin{\theta_{ij}}$ (for \textit{i,j = 1,2,3}),  and $P_{m}=diag(e^{i\rho_{1}},e^{i\rho_{2}},0)$ denotes the diagonal Majorana phase matrix, which does not have any effect on the neutrino oscillations.

Neutrino oscillation is typically measured by comparing the flux of produced $\alpha$-flavor neutrinos and the flux of $\beta$-flavor neutrinos observed in a detector placed at some distance from the production source. The probability for an $\alpha$-flavor to oscillate into a $\beta$-flavor, $P_{({\nu_{\alpha}}\rightarrow\nu_{\beta})}$, depends on three mixing angles ($\theta_{12},\theta_{13},\theta_{23}$), the \textit{CP}-violating phase (\dcp), two mass-squared splittings ($\Delta m^{2}_{21}$, $\Delta m^{2}_{31}$) where $\Delta m^2_{ij}=m^2_{i}-m^2_{j}$, its energy $(E_{\nu})$, the propagation distance $(L)$, and the density of the matter passed through by the neutrino $\rho$, given by
\begin{equation*}
    P_{({\nu_{\alpha}}\rightarrow\nu_{\beta})}=f\left(\theta_{12},\theta_{13},\theta_{23},\delta_{\text{CP}};\Delta m^{2}_{21}, \Delta m^{2}_{31};E_{\nu},L,\rho\right).
\end{equation*}

It is well-established from the contribution of many neutrino experiments~\cite{Zyla:2020zbs}, using both the natural neutrino sources (solar and atmospheric neutrinos) and the man-made neutrino sources (reactor and accelerator neutrinos) that the two leptonic mixing angles, $\theta_{12}$ and $\theta_{23}$, are large, $\theta_{13}$ is relatively small but nonzero, and the mass-squared splitting $|\Delta m^{2}_{31}|$ is about 30 times larger than $\Delta m^{2}_{21}$. The global analysis of neutrino oscillation data is available, e.g., in Ref.~\cite{esteban2019global,Esteban:2020cvm}, and is briefly summarized in Table~\ref{tab:nuoscpara}.
\begin{table}
    \centering
    \begin{tabular}{l|r}
    \hline\hline
    Parameter & Best fit$\pm1\sigma$\\\hline
    $\sin^{2}\theta_{12}$ & $0.310^{+0.013}_{-0.012}$\\ $\sin^{2}\theta_{13}(\times 10^{-2})$ & $2.241^{+0.067}_{-0.066}$\\ $\sin^{2}\theta_{23}$ & $0.558^{+0.020}_{-0.033}$\\ $\delta_{\text{CP}}(^{\circ})$ & $222^{+38}_{-28}$\\
    $\Delta m^{2}_{21} (10^{-5}\text{eV}^{2}/c^4)$ & $7.39^{+0.21}_{-0.20}$\\
    $\Delta m^{2}_{31} (10^{-3}\text{eV}^{2}/c^{4})$ & $2.523^{+0.032}_{-0.030}$ \\\hline
    \end{tabular}
    \caption{Global constraint of oscillation parameters with \emph{normal} mass hierarchy assumed, taken from  Refs.~\cite{esteban2019global,nufit41}}.
    \label{tab:nuoscpara}
\end{table}
\noindent Although a few percent precision measurements of three mixing angles and two mass-squared splittings have been achieved, a complete picture of neutrino oscillation has not been fulfilled yet. There are at least three \emph{unknowns}, which the worldwide neutrino programs plan to address in the next decades. The first \emph{unknown} is \textit{CP} violation (CPV) in the neutrino oscillations. Despite a recent hint of maximal CPV from the \dcp\ measurement by the T2K experiment~\cite{Abe:2019vii}, whether \textit{CP} is violated or not requires higher statistics to be established. The second \emph{unknown} is the neutrino mass hierarchy (MH), which refers to the order of the three mass eigenvalues of neutrino mass eigenstates. Whether the MH is \emph{normal}  ($m_{1}<m_{2}<m_{3}$) or \emph{inverted} ($m_{3}<m_{1}<m_{2}$) is still questionable. While the recent measurements from individual experiments~\cite{t2knew,NOvAnew,minosnew} mildly favor the former, the efforts~\cite{Esteban:2020cvm,Kelly:2020fkv} for fitting jointly multiple neutrino data samples show that the preference to the normal MH becomes less significant. Thus, more neutrino data is essential to shedding light on the neutrino MH. The third \emph{unknown} on the list is about the mixing angle \thetamu.  Its measured value is close to $45^\circ$, which means the mass eigenstate $\nu_3$ is comprised of an approximately equal amount of $\nu_{\mu}$ and $\nu_{\tau}$, indicating some \emph{unknown} symmetry between the second and third lepton generations. Whether \thetamu\ is exactly equal to $45^\circ$ in the lower octant (LO, $\theta_{23}<45^\circ$) or in the higher octant (HO, $\theta_{23}>45^\circ$) is of interest to pursue.

In this paper, we show the prospect of reaching these \emph{unknowns} in light of two accelerator-based long-baseline neutrino experiments, T2K-II and \nova\ extended program, and a reactor-based medium-baseline neutrino experiment, JUNO. The paper is organized as follows. Section~\ref{sec:simulate} details the experimental specifications of these three experiments and elaborates on the simulation methodology. In Sec.~\ref{sec:results}, we present our results on the MH determination, the CPV sensitivity, the resolution of the \thetamu\ octant, and the precise constraints of the oscillation parameters. We give the conclusion of the work in Sec.~\ref{sec:fin}.

\section{\label{sec:simulate}Experimental specifications and simulation details}
\subsection{Experimental specifications of T2K-II, NO$\nu$A-II and JUNO}
 \textbf{T2K-II}: The ongoing Tokai-To-Kamioka (T2K)~\cite{Abe:2011ks} is the second generation of accelerator-based long-baseline (A-LBL) neutrino oscillation experiments located in Japan, and T2K-II~\cite{abe2016sensitivity} is a proposal to extend the T2K run until 2026 before Hyper-Kamiokande~\cite{Abe:2018uyc} starts operation. The T2K far detector, SK, is located 295 km away from the neutrino production source, and receives the neutrino beam at an average angle of $2.5^o$ off-axis to achieve a narrow-band neutrino beam with a peak energy of 0.6 GeV. Being a gigantic Cherenkov detector with 50~ktons of pure water and approximately 13,000 photomultiplier tubes deployed, SK provides an excellent performance of reconstructing the neutrino energy and the neutrino flavor classification. This capability allows T2K(-II) to measure simultaneously the disappearance of muon (anti-)neutrinos and the appearance of electron (anti-)neutrinos from the flux of almost pure muon (anti-)neutrinos. While the data samples of the \numu\ (\numub) disappearance provide a precise measurement of the atmospheric neutrino parameters, $\sin^22\theta_{23}$ and \dmatm, the \nue\ (\nueb) appearance rates are driven by $\sin^22\theta_{13}$ and are sensitive to \dcp\ and the MH. The sensitivity of the A-LBL experiments such as T2K and \nova\ to \dcp\ and the MH can be understood via the following expression of the so-called \textit{CP} asymmetry~\cite{Suekane:2015yta}, presenting a relative difference between $P_{(\nu_{\mu}\rightarrow \nu_e)}$ and $P_{(\bar{\nu}_{\mu}\rightarrow \bar{\nu}_e)}$ near the oscillation maximum, and corresponding to $\frac{|\Delta m^2_{31}|L}{4E_{\nu}}=\pi/2$. 

 \begin{align} \label{eq:adcp} 
    A_{\text{CP}}&\left(\frac{|\Delta m^2_{31}|L}{4E_{\nu}}=\pi/2\right) = \frac{P_{(\nu_{\mu}\rightarrow \nu_e)}- P_{(\bar{\nu}_{\mu}\rightarrow \bar{\nu}_e)}}{P_{(\nu_{\mu}\rightarrow \nu_e)} + P_{(\bar{\nu}_{\mu}\rightarrow \bar{\nu}_e)}} \nonumber\\ 
    &\sim -\frac{\pi \sin 2\theta_{12}}{\tan\theta_{23}\sin 2\theta_{13}} \frac{\Delta m^2_{21}}{|\Delta m^2_{31}|} \sin \delta_{\text{CP}} \pm \frac{L}{2800km},
\end{align}
 
 \noindent where the $+(-)$ sign is taken for the \emph{normal} (\emph{inverted}) MH, respectively. With the values listed in Table~\ref{tab:nuoscpara}, $\frac{\pi \sin 2\theta_{12}}{\tan\theta_{23}\sin 2\theta_{13}} \frac{\Delta m^2_{21}}{|\Delta m^2_{31}|} \sim 0.256$, which means the \textit{CP} violation effect can be observed somewhat between $-25.6\%$ and $+25.6\%$. For a 295~km baseline of the T2K experiment, the mass hierarchy effect is subdominant with $\sim 10.5\%$. T2K uses a near detector complex, situated 280 m from the production target to constrain the neutrino flux and the neutrino interaction model. T2K made an observation of electron neutrinos appearing from a muon neutrino beam~\cite{Abe:2013hdq} and presented an indication of CPV in the neutrino oscillation~\cite{Abe:2019vii}. T2K originally planned to take data equivalent to $7.8\times 10^{21}$ protons-on-target (POT) exposure. At the Neutrino 2020 conference, T2K~\cite{patrick_dunne_2020_4154355} reported a collected data sample from $3.6\times 10^{21}$ POT exposure. In Ref.~\cite{abe2016sensitivity}, T2K proposes to extend the run until 2026 to collect $20\times 10^{21}$ POT, allowing T2K to explore CPV with a confidence level (\conlev) of $3\sigma$ or higher if \dcp\ is close to $-\pi/2$ and to make precision measurements of \thetamu\ and \absdmatm.

\textbf{\nova\ extension or \nova-II: } Ongoing NuMI Off-axis $\nu_e$ Appearance (\nova\ )~\cite{ayres2007nova} is also the second generation of A-LBL neutrino experiments placed in the United States with a baseline of 810 km between the production source and the far detector. Such a long baseline allows \nova\ to explore the MH with high sensitivity via the matter effect~\cite{Wolfenstein:1977ue} on the (anti-)neutrino interactions. From Eq.~(\ref{eq:adcp}), it can be estimated that the matter effect in \nova\ is $\sim 28.9\%$, which is slightly higher than the \textit{CP} violation effect. However, these two effects, along with the ambiguity of the $\theta_{23}$ octant, are largely entangled. In other words, \nova\ sensitivity on the neutrino MH depends on the value of \dcp. \nova's recent data~\cite{alex_himmel_2020_4142045} does not provide as much preference to the neutrino mass hierarchy as T2K~\cite{patrick_dunne_2020_4154355} does since \nova\ data shows no indication of the \textit{CP} violation. Similar to T2K, \nova\ adopts the off-axis technique such that the far detector is placed at an angle of 14~mrad to the averaged direction of the neutrino beam. \nova\ uses a near detector, located 1~km away from the production target, to characterize the unoscillated neutrino flux. The \nova\ far detector is filled with liquid scintillator contained in PVC cells, totally weighted at 14 ktons with 63$\%$ active materials. \nova\ takes advantage of machine learning for particle classification to enhance the event selection performance. In 2018~\cite{acero2019first}, \nova\ provided more than a 4$\sigma$ \conlev\ evidence of electron anti-neutrino appearance from a beam of muon anti-neutrinos. At the Neutrino 2020 conference, \nova~\cite{alex_himmel_2020_4142045} reported a collected data sample from $2.6\times 10^{21}$ POT exposure. In~\cite{sanchez2018nova}, \nova\ gives the prospect of extending the run through 2024, hereby called \nova-II, in order to get a 3$\sigma$ \conlev\ or higher sensitivity to the MH in case the MH is \emph{normal} and \dcp\ is close to $-\pi/2$, and more than a 2$\sigma$ \conlev\ sensitivity to CPV.

\textbf{JUNO:} Jiangmen Underground Neutrino Observatory (JUNO)~\cite{djurcic2015juno} is a reactor-based medium-baseline neutrino experiment located in China. JUNO houses a 20~kton large liquid scintillator detector for detecting the electron anti-neutrinos ($\overline{\nu}_e$) from the Yangjiang (YJ) and Taishan (TS) nuclear power plants (NPPs) with an average baseline of 52.5~km. Each of the six cores at the YJ nuclear plant will produce a power of 2.9~GW and the four cores at the TS NPP will generate 4.6~GW each. They are combined to give 36~GW of thermal power. JUNO primarily aims to determine the MH by measuring the surviving $\overline{\nu}_e$ spectrum, which uniquely displays the oscillation patterns driven by both solar and atmospheric neutrino mass-squared splittings~\cite{Zhan:2008id}. This feature can be understood via the $\bar{\nu}_e$ disappearance probability in the vacuum, which is expressed as follow:

\begin{align}\label{eq:nuebdis}
    &P_{(\bar{\nu}_e\rightarrow\bar{\nu}_e)} = 1-\cos^4\theta_{13}\sin^22\theta_{12}\sin^2\Phi_{21}\nonumber\\
    &-\sin^22\theta_{13}\left( \cos^2\theta_{12}\sin^2\Phi_{31} +\sin^2\theta_{12}\sin^2\Phi_{32} \right),
\end{align}

\noindent where $\Phi_{ij}=\frac{\Delta m^2_{ij}L}{4E_{\nu}}$. An averaged 52~km baseline of the JUNO experiment obtains the maximum oscillation corresponding to $\Phi_{21}=\pi/2$ around 3~MeV, and relatively enhances the oscillation patterns driven by the $\Phi_{31}$ and $\Phi_{32}$ terms. The relatively small difference between $\Delta m^2_{31}$ and $\Delta m^2_{32}$ make oscillation patterns in the \emph{normal} and \emph{inverted} MH scenarios distinguishable. To realize practically the capability of mass hierarchy resolution, JUNO must achieve a very good neutrino energy resolution, which has been demonstrated recently in Ref.~\cite{Abusleme:2020lur}, and collect a huge amount of data. With six years of operation, JUNO can reach a 3$\sigma$ \conlev\ or higher sensitivity to the MH and achieve better than $1\%$ precision on the solar neutrino parameters and the atmospheric neutrino mass-squared splitting \absdmatm.

Although T2K and \nova\ experiments have already collected $18\%$ and $36\%$ of the total proton exposure assumed in this study, respectively, we do not directly use their experimental data to estimate their final reaches. The main reason is that measurements of the \textit{CP} violation, the mass hierarchy, and the mixing angle $\theta_{23}$ are so far statistically limited except for a specific set of oscillation parameters. We thus carry out the study with the assumption that all values of \dcp\ and the two scenarios of the neutrino mass hierarchy are still possible, and the mixing angle $\theta_{23}$ is explored in a range close to $45^{\circ}$. 
 
Reaching the three above-mentioned \emph{unknowns} depends on the ability to resolve the parameter degeneracies among \dcp, the sign of \dmatm, $\theta_{13}$, and \thetamu~\cite{Barger:2001yr}. Combining the data samples of the A-LBL experiments (T2K-II and \nova-II) and JUNO would enhance the CPV search and the MH determination since the JUNO sensitivity to the MH has no ambiguity to \dcp. To further enhance the CPV search, one can break the \dcp-$\theta_{13}$ degeneracy by using the constraint of $\theta_{13}$ from reactor-based short-baseline (R-SBL) neutrino experiments such as Daya Bay~\cite{Guo:2007ug}, Double Chooz~\cite{Ardellier:2006mn}, and RENO~\cite{Ahn:2010vy}. This combination also helps to solve the \thetamu\ octant in the case of nonmaximal mixing.

\begin{table*}
    \caption{\label{tab:lbl}  Experimental specifications of the A-LBL experiments T2K-II and NO$\nu$A-II}
    \begin{ruledtabular}
    \begin{tabular}{lll}
    \textbf{Characteristics} & \textbf{T2K-II}~\cite{abe2016sensitivity,Abe:2017vif} & \textbf{NO$\nu$A-II}~\cite{sanchez2018nova,acero2019first}  \\\hline
    \textbf{Baseline} & 295 km & 810 km\\
    \textbf{Matter density}~\cite{dziewonski1981preliminary} & 2.6 $gcc^{-1}$ & 2.84 $gcc^{-1}$\\
    \textbf{Total exposure} & $20\times10^{21}$ POT~ & $72\times10^{20}$ POT\\
    \textbf{Detector fiducial mass} & 22.5 kton & 14 kton\\
    \textbf{Systematics\footnote{Normalization (calibration) error for both signals and backgrounds.}} & 3\% (0.01\%) & 5\% (2.5\%)\\
    \textbf{Energy resolution} & $0.03\times \sqrt{\text{E(GeV)}}$ & $x\times \sqrt{\text{E(GeV)}}$\footnote{$x=0.107$, $0.091$, $0.088$, and $0.081$ for $\nu_e$, $\nu_\mu$, $\bar{\nu}_e$, and $\bar{\nu}_\mu$, respectively.}\\
   \textbf{Energy window} & 0.1-1.3 GeV (\textit{APP}\footnote{Shortened for the appearance sample.}), 0.2-5.05 GeV (\textit{DIS}\footnote{Shortened for the disappearance sample.}) & 0.0-4.0 GeV (\textit{APP}), 0.0-5.0 GeV (\textit{DIS})\\
   \textbf{Bin width} & 0.125 GeV/bin (\textit{APP}), variable\footnote{Used the binning as in \cite{acero2019first}.} (\textit{DIS}) & 0.5 GeV/bin (\textit{APP}), variable (\textit{DIS})\\
    \end{tabular}
    \end{ruledtabular}
\end{table*}
\begin{table*}
    \caption{\label{tab:lblapp}  Detection efficiencies\footnote{Defined per each interaction channel as the ratio of selected events in the data sample to the totally simulated interaction supposed to happen in the detector.}(\%) of signal and background events in appearance samples. Normal mass hierarchy and $\delta_{CP}=0$ are assumed.}
    \begin{ruledtabular}
    \begin{tabular}{l|c|ccccccc}
     & & $\nu_\mu \rightarrow\nu_e$ & $\bar{\nu}_\mu \rightarrow \bar{\nu}_e$ & $\nu_\mu$ CC & $\bar{\nu}_\mu$ CC & $\nu_e$ CC & $\bar{\nu}_e$ CC & NC\\\hline
     \textbf{T2K-II} & $\nu$ mode & 65.5 & 46.2 & 0.02 & 0.02 & 19.8 & 19.8 & 0.41\\
     & $\bar{\nu}$ mode & 45.8 & 70.7 & 0.01 & 0.01 & 17.5 & 17.5 & 0.45\\\hline
     \textbf{NO$\nu$A-II} & $\nu$ mode & 62.0 & 38.0 & 0.15 & -- & 79.0 & 69.0 & 0.87\\
     & $\bar{\nu}$ mode & 25.0 & 67.0 & 0.14 & 0.05 & 20.7 & 40.7 & 0.51\\
    \end{tabular}
    \end{ruledtabular}
\end{table*}

\begin{table*}
    \caption{\label{tab:lbldisapp}  Detection efficiencies(\%) of signal and background events in disappearance samples. Normal mass hierarchy is assumed.}
    \begin{ruledtabular}
    \begin{tabular}{l|c|cc|cc|ccc}
     & & $\nu_\mu$ CCQE  & $\nu_\mu$ CC non-QE & $\bar{\nu}_\mu$ CCQE & $\bar{\nu}_\mu$ CC non-QE & ($\nu_e$ + $\bar{\nu}_e$) CC &  NC & $\nu_\mu \rightarrow\nu_e$\\\hline
     \textbf{T2K-II} & $\nu$ mode & 71.2 & 20.4 & 71.8 & 20.4 & 0.84 & 2.7 & 0.84\\
     & $\bar{\nu}$ mode & 65.8 & 24.5 & 77.5 & 24.5 & 0.58 & 2.5 & 0.58\\\hline
     \textbf{NO$\nu$A-II} & $\nu$ mode & \multicolumn{2}{c}{31.2\footnote{The efficiency for CCQE and CC non-QE interactions are considered equal.} }& \multicolumn{2}{c}{27.2} & -- & 0.44 & -- \\
     & $\bar{\nu}$ mode & \multicolumn{2}{c}{33.9}& \multicolumn{2}{c}{20.5} & -- & 0.33 & --\\
    \end{tabular}
    \end{ruledtabular}
\end{table*}

\subsection{Simulation details} 
The General Long-Baseline Experiment Simulator (GLoBES)~\cite{Huber:2004ka,huber2007new} is used for simulating the experiments and calculating their statistical significance. In this simulator, a number of expected events of $\nu_j$ from $\nu_i$ oscillation in the \textit{n}-th energy bin of the detector in a given experiment is calculated as
\begin{align}
    &R_{n}(\nu_{i} \rightarrow \nu_{j}) = \frac{N}{L^{2}} \int_{E_{n}-\Delta E_{n}}^{E_{n}+\Delta E_{n}}dE_{r}\times \nonumber\\
    &\int dE_{t}\Phi_{i}(E_{t})\sigma_{\nu_{j}} R_{j}(E_{t},E_{r})\epsilon_{j}(E_{r}) P_{\nu_{i}\rightarrow\nu_{j}}(E_{t})
\end{align}
where \textit{i}, \textit{j} are the charged lepton(s) associated with the initial and final flavor(s) of the neutrinos, $\Phi_{i}$ is the flux of the initial flavor at the source, $\sigma_{\nu_{j}}$ is the cross-section for the final flavor \textit{f}, \textit{L} is the baseline length, $E_{t}$ and $E_{r}$ are the incident and reconstructed neutrino energy, respectively, $\epsilon_{j}(E_{r})$ is the detection efficiency of the final flavor \textit{f}, and \textit{N} is the normalization factor for standard units in GLoBES. We describe the experiments using updated information of fluxes, signal and background efficiencies, and systematic errors. Remaining differences between the energy spectra of the simulated data sample at the reconstruction level obtained by GLoBES and the real experiment simulation can be due to the effects of the neutrino interaction model, the detector acceptance, detection efficiency variation as a function of energy, etc... These differences are then treated quantitatively using post-smearing efficiencies, consequently allowing us to match our simulation with the published spectra of each simulated sample from each experiment. Each experimental setup is validated at the event rate level and sensitivity level to ensure that physics reaches of the simulated data samples we obtain are in relatively good agreement with the real experimental setup. \\
For each T2K-II and \nova-II experiment, four simulated data samples per each experiment are used: $\nu_{\mu}(\bar{\nu}_{\mu})$ disappearance and $\nu_{e}(\bar{\nu}_{e})$ appearance in both \numode\ and \antinumode. The experimental specifications of these two experiments are shown in Table~\ref{tab:lbl}.
\begin{figure*}
    \includegraphics[scale=0.4]{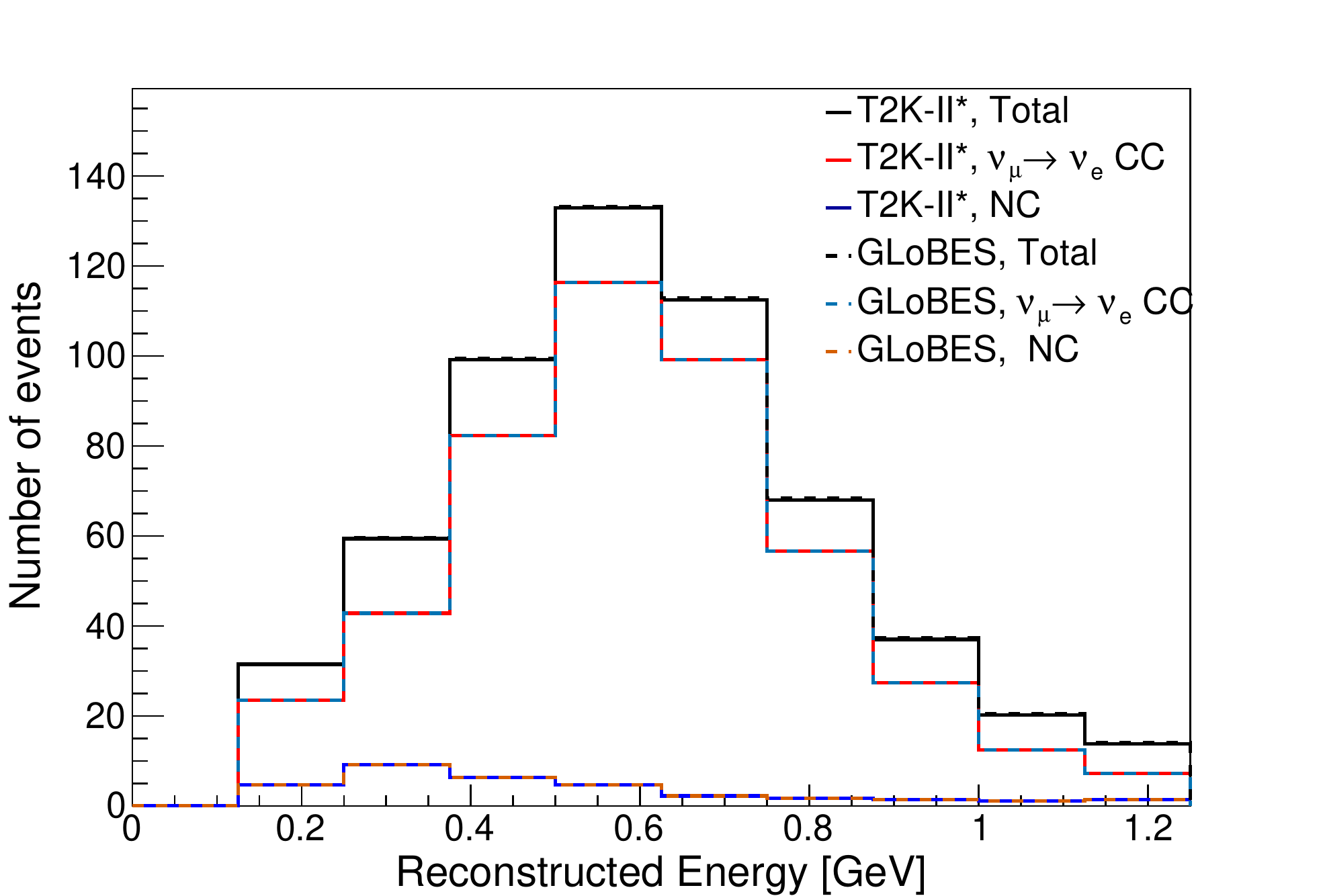}
    \includegraphics[scale=0.4]{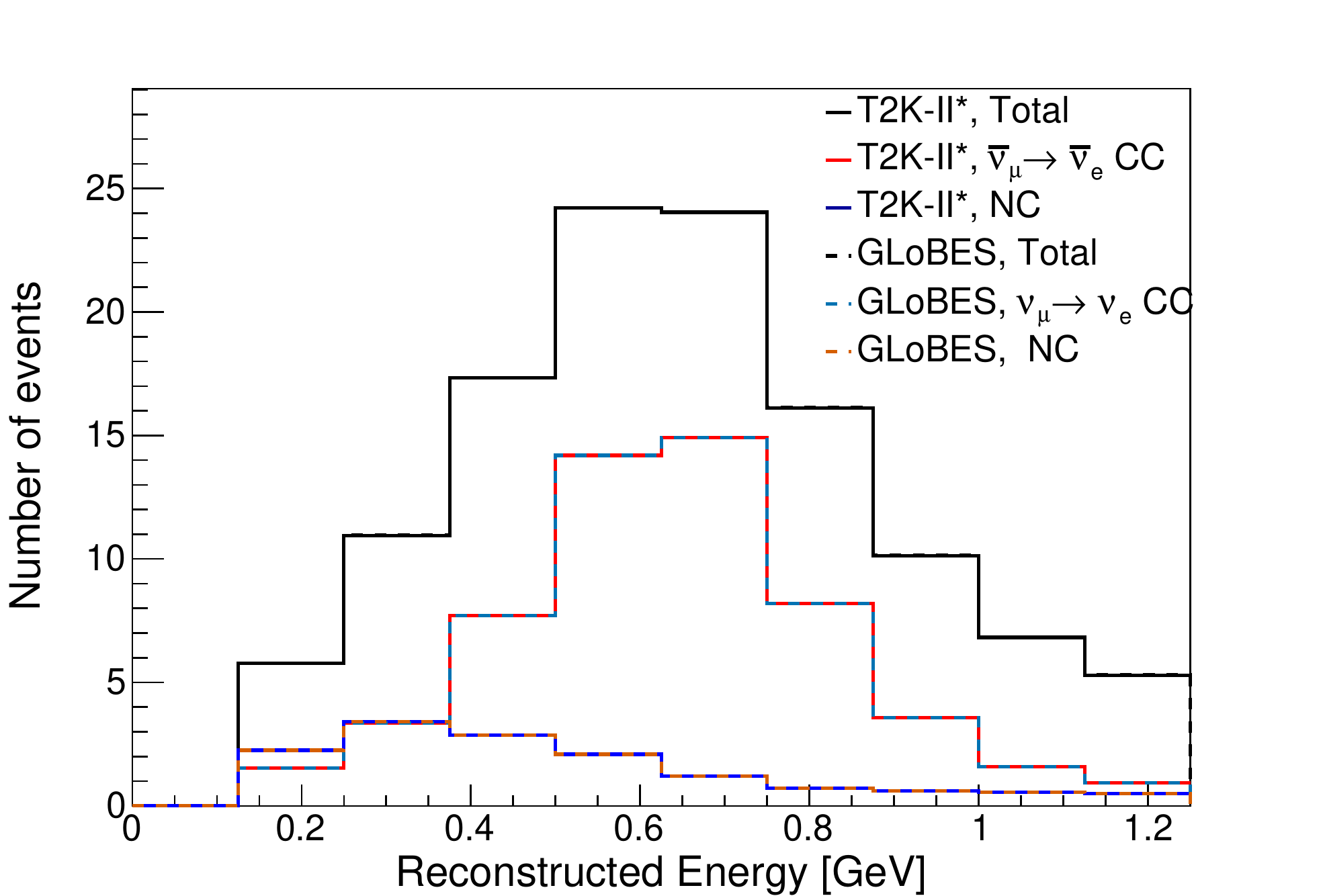}
    \includegraphics[scale=0.4]{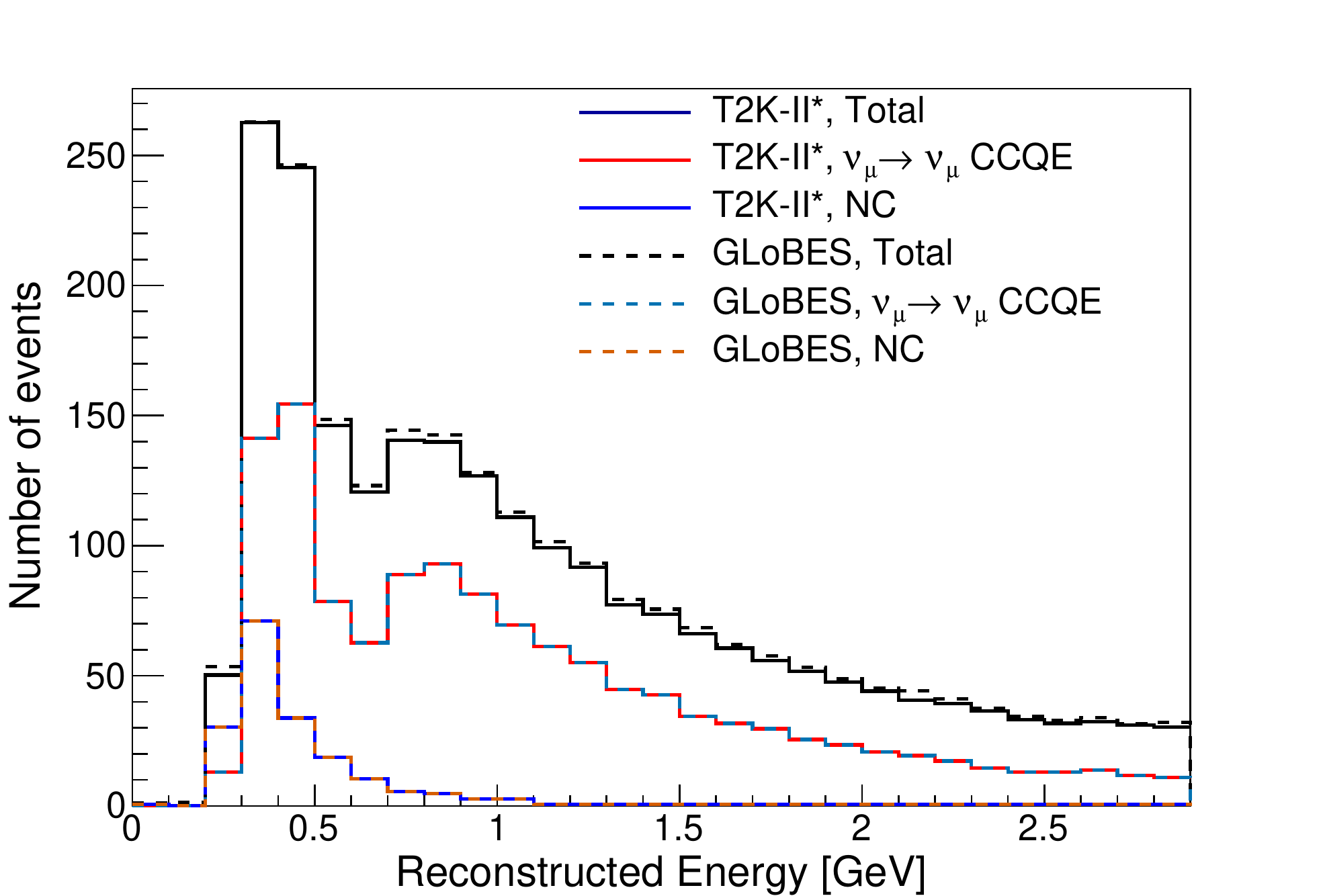}
    \includegraphics[scale=0.4]{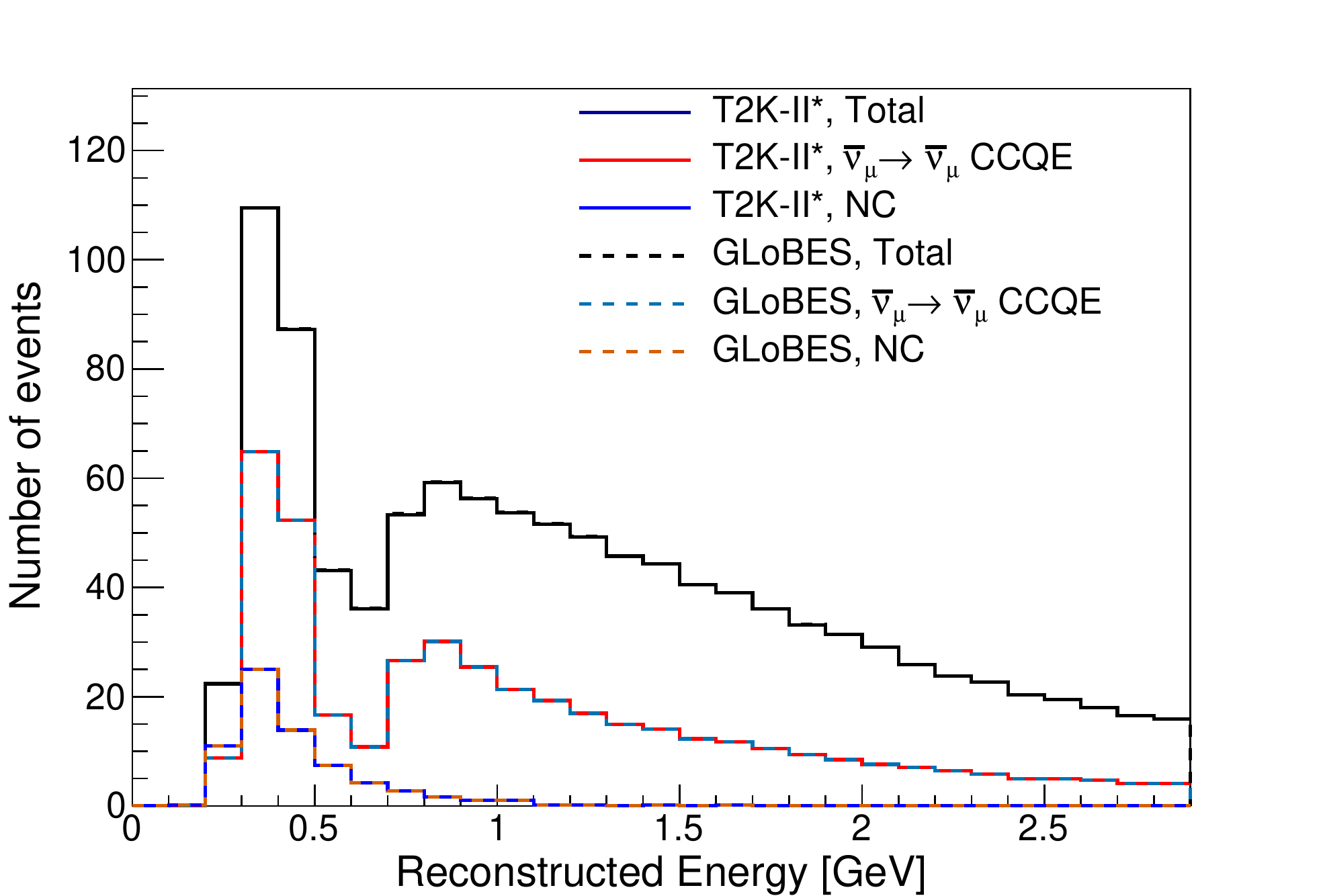}
    \caption{\label{fig:t2kspectra}Expected event spectra of the signal and background as a function of reconstructed neutrino energy for T2K-II. The top (bottom) spectra are for the appearance (disappearance) samples and the left (right) spectra are for \numode\ (\antinumode). The same oscillation parameters as in Ref.~\cite{Abe:2017vif} are used.}
\end{figure*}
In T2K(-II), neutrino events are dominated by the charged current quasielastic (CCQE) interactions. Thus, for appearance (disappearance) in \numode\ and \antinumode, the signal events are obtained from the $\nu_{\mu}\rightarrow\nu_{e}$ ($\nu_{\mu}\rightarrow\nu_{\mu}$) CCQE events and the $\bar{\nu}_{\mu}\rightarrow\bar{\nu}_{e}$ ($\bar{\nu}_{\mu}\rightarrow\bar{\nu}_{\mu}$) CCQE events, respectively. In the appearance samples, the intrinsic $\nu_{e}/\bar{\nu}_{e}$ contamination from the beam, the \emph{wrong-sign} components, i.e., $\overline{\nu}_{\mu}\rightarrow\overline{\nu}_{e}$ ($\nu_{\mu}\rightarrow\nu_{e}$) in \numode\ (\antinumode), respectively, and the neutral current (NC) events constitute the backgrounds. In the disappearance samples, the backgrounds come from $\nu_\mu$, $\overline{\nu}_\mu$ charged current (CC) interaction excluding CCQE, hereby called CC non-QE, and NC interactions. We use the updated T2K flux released along with Ref.~\cite{Abe:2015awa}. In simulation, the cross section for low- and high-energy regions are taken from Refs.~\cite{Messier:1999kj,Paschos:2001np}, respectively. In our T2K-II setup, an exposure of $20\times10^{21}$ POT equally divided among the \numode\ and the \antinumode\ is considered, along with a 50\% effectively statistic improvement as presented in Ref.~\cite{abe2016sensitivity}. The signal and background efficiencies and the spectral information for T2K-II are obtained by scaling the T2K analysis reported in Ref.~\cite{Abe:2017vif} to the same exposure as the T2K-II proposal. In Fig.~\ref{fig:t2kspectra}, the T2K-II expected spectra of the signal and background events as a function of reconstructed neutrino energy obtained with GLoBES are compared to those of the Monte-Carlo simulation scaled from Ref.~\cite{abe2016sensitivity}. A $3\%$ error is assigned for both the energy resolution and the normalization uncertainties of the signal and background in all simulated samples.
\begin{figure*}
    \centering
    \includegraphics[scale=0.405]{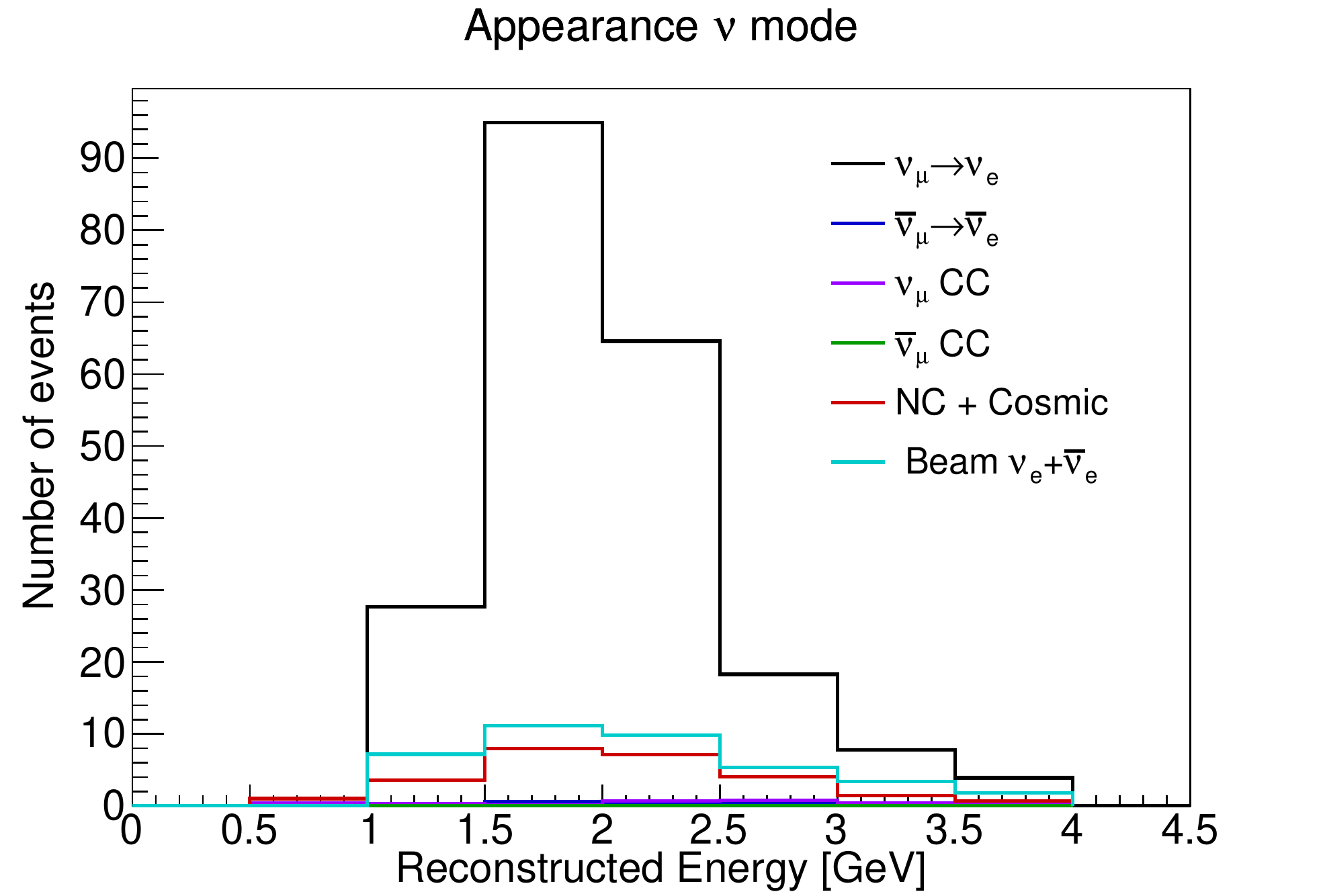}
    \includegraphics[scale=0.405]{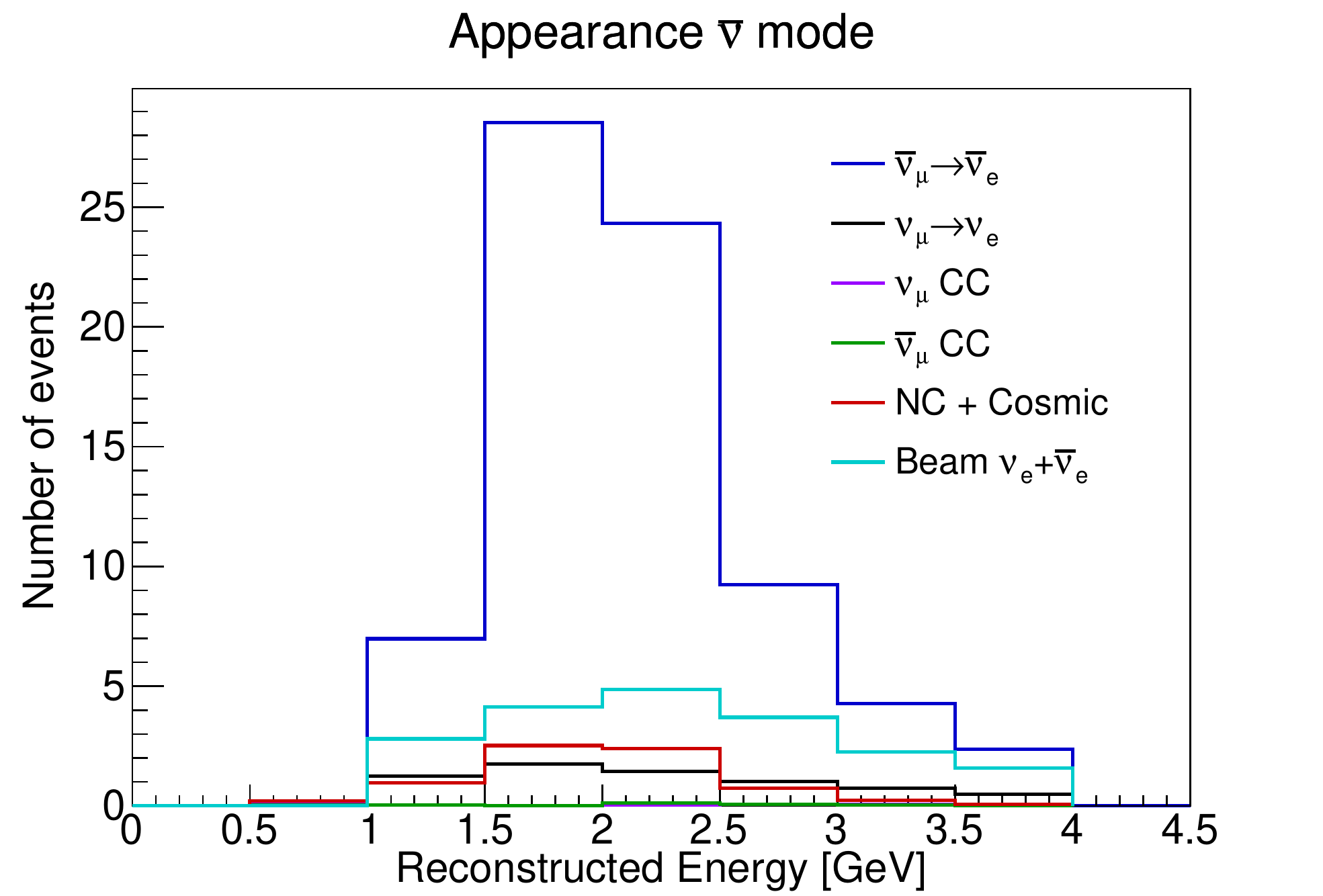}
    \includegraphics[scale=0.405]{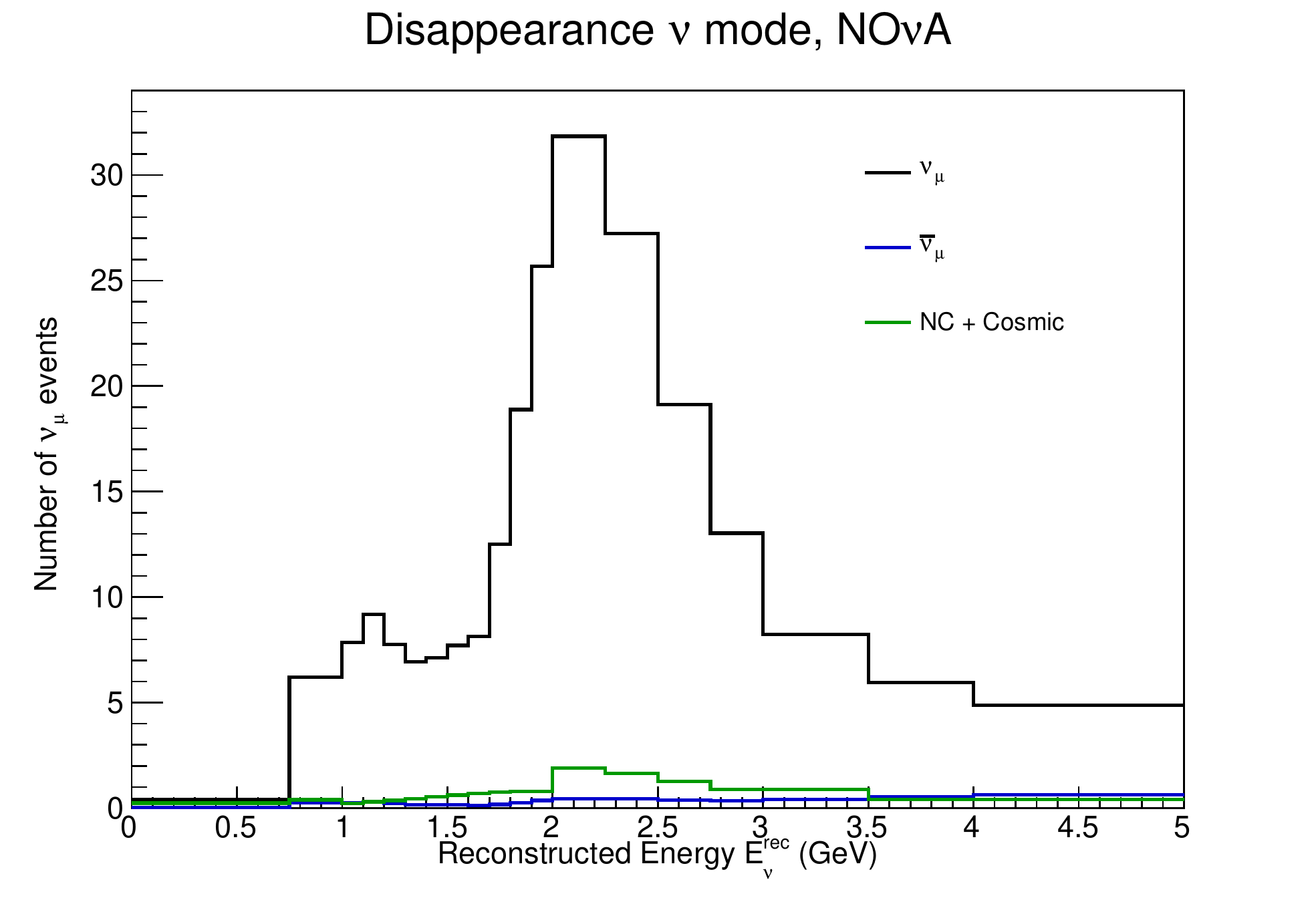}
    \includegraphics[scale=0.405]{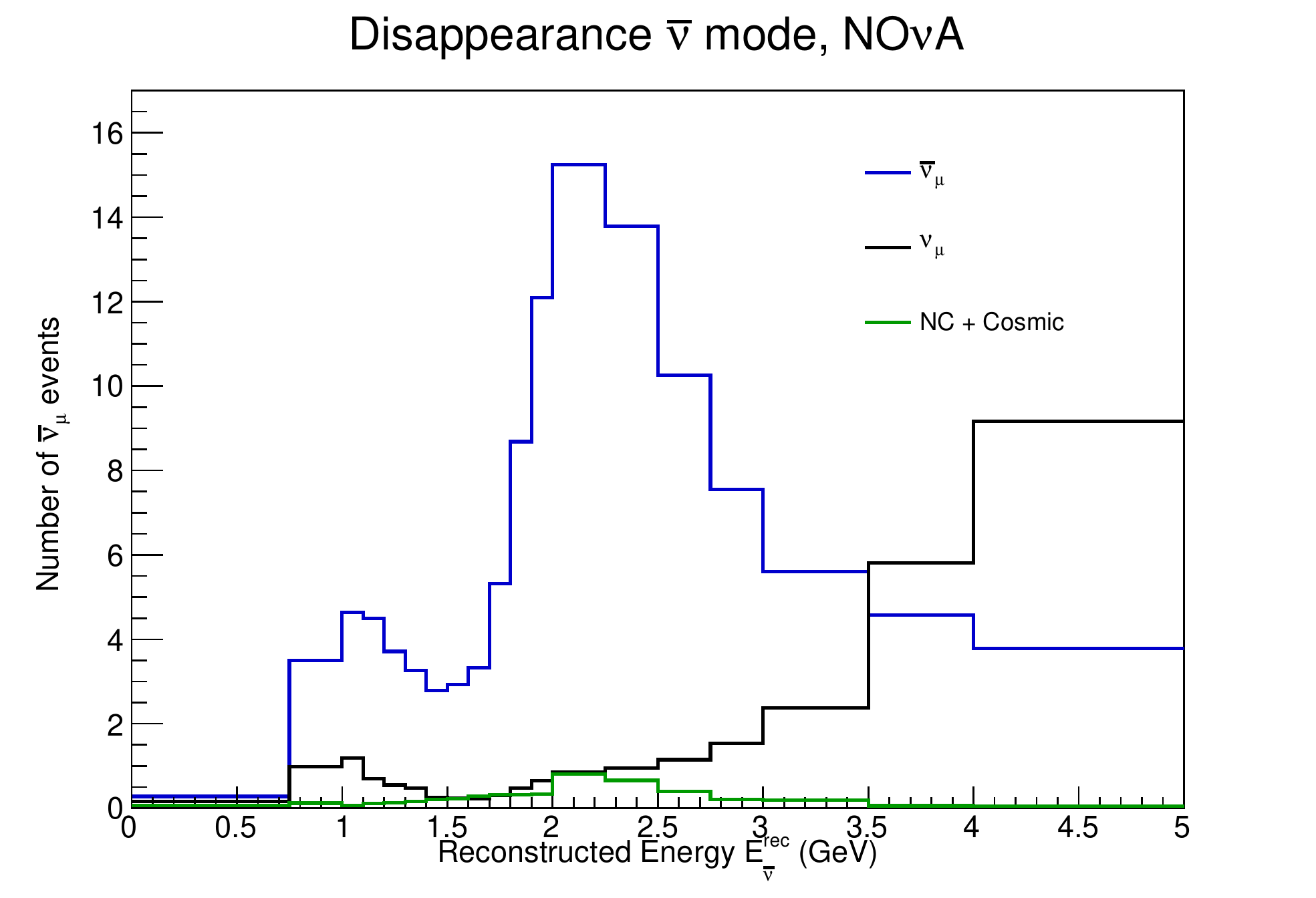}
    \caption{\label{fig:novaspectra}Expected event spectra of the signal and background as a function of reconstructed neutrino energy for \nova-II. The top (bottom) spectrum is for the appearance (disappearance) channel and the left (right) spectrum is for \numode\ (\antinumode). \emph{Normal} MH, $\delta_{\text{CP}}=0$, and other oscillation parameters given in Table~\ref{tab:nuoscpara}  are assumed.}
\end{figure*}

For \nova-II, we consider a total exposure of $72\times10^{20}$ POT equally divided among \numode\ and \antinumode~\cite{sanchez2018nova}. We predict the neutrino fluxes at the \nova\ far detector by using the flux information from the near detector, given in Ref.~\cite{NDfluxnova}, and normalizing it with the square of their baseline ratio.  A 5\% systematic error for all samples and 8\textup{--}10\% sample-dependent energy resolutions are assigned. Significant background events in the appearance samples stem from the intrinsic beam $\nu_{e}/\bar{\nu}_{e}$, NC components, and cosmic muons. In the appearance sample of the \antinumode, \emph{wrong-sign} events from $\nu_e$ appearance events are included as the backgrounds in the simulation. We use the reconstructed energy spectra of the \nova\ far detector simulated sample, reported in Ref.~\cite{acero2018new}, to tune our GLoBES simulation. The low- and high-particle identification score samples are used, but the peripheral sample is not since the reconstructed energy information is not available. In the disappearance samples of both \numode\ and \antinumode, events from both CC $\nu_\mu$ and $\bar{\nu}_\mu$ interactions are considered to be signal events, which is tuned to match with the \nova\ far detector simulated signal given an identical exposure. Background from the NC $\nu_\mu$ ($\bar{\nu}_\mu$) interactions is taken into consideration and weighted such that the rate at a predefined exposure is matched to a combination of the reported NC and cosmic muon backgrounds in Ref.~\cite{acero2018new}.
\noindent Figure~\ref{fig:novaspectra} shows the simulated \nova-II event spectra as a function of reconstructed neutrino energy for $\nu_e$ appearance and $\nu_\mu$ disappearance channels in both \numode\ and \antinumode, where \emph{normal} MH is assumed, $\delta_{\text{CP}}$ is fixed at $0^\circ$, and other parameters are given in Table~\ref{tab:nuoscpara}.

Tables~\ref{tab:lblapp} and~\ref{tab:lbldisapp} detail our calculated signal and background detection efficiencies for the electron (anti-)neutrino appearance and muon (anti-)neutrino disappearance, respectively, in T2K and \nova. The two neutrino experiments reach a relatively similar performance for selecting the electron (anti-)neutrino appearance samples. While T2K gains due to the excellent separation of muons and electrons with the water Cherenkov detector, \nova\ boosts the selection performance with the striking features of the liquid scintillator and the powerful deep learning. For selecting the disappearance samples, T2K outperforms since the T2K far detector is placed deep underground while the NOvA far detector is on the surface and suffers a much higher rate of cosmic ray muons.

\begin{figure}
\centering
\includegraphics[width=.9\linewidth]{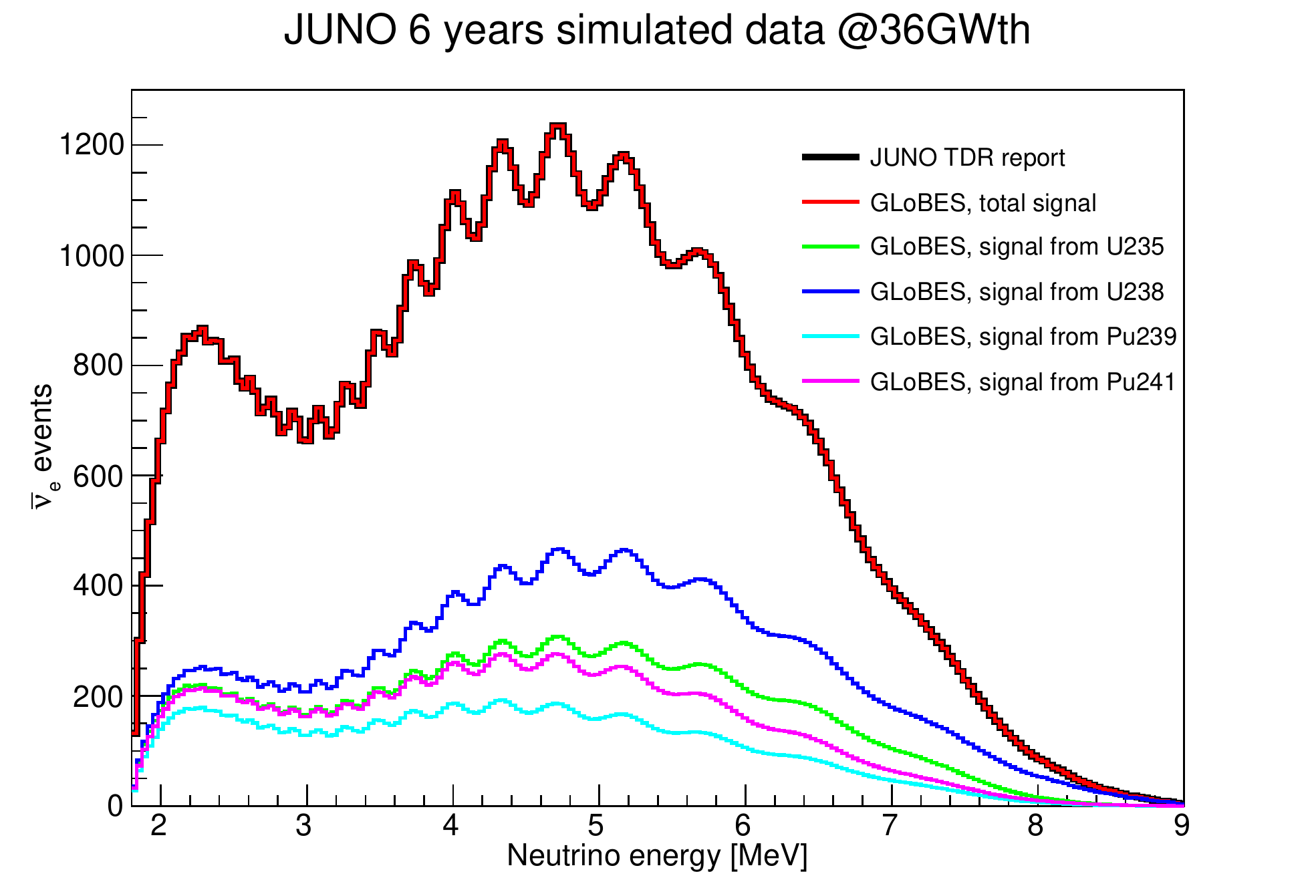}
\caption{\label{fig:junospectra}JUNO event rate calculated at the same oscillation parameters as Ref.~\cite{djurcic2015juno}}
\end{figure}

In JUNO, the electron anti-neutrino \nueb\ flux, which is produced mainly from four radioactive isotopes ($^{235}\text{U}$, $^{238}\text{U}$, $^{239}\text{Pu}$, and $^{241}\text{Pu}$~\cite{huber2004precision}), is simulated with an assumed detection efficiency of $73\%$. The backgrounds, which have a marginal effect on the MH sensitivity, are not included in our simulation. In our setup, to speed up the calculation, we consider one core of 36~GW thermal power with an average baseline of 52.5~km instead of the true distribution of the reactor cores, baselines, and powers. The simulated JUNO specification is listed in Table~\ref{tab:juno} and the event rate distribution as a function of the neutrino energy is shown in Fig.~\ref{fig:junospectra}. For systematic errors, we commonly use $1\%$ for the errors associated with the uncertainties of the normalization of the \nueb\ flux produced from the reactor core, the normalization of the detector mass, the spectral normalization of the signal, the detector response to the energy scale, the isotopic abundance, and the bin-to-bin reconstructed energy shape.
\begin{table}
    \begin{ruledtabular}
    \begin{tabular}{c@{\hskip 0.in}@{\hskip 0.in}c}
    \textbf{Characteristics} & \textbf{Inputs}\\\hline
    Baseline & 52.5 km\\
    Density & 2.8 $gcc^{-1}$~\cite{khan2020matter}\\
    Detector type & Liquid Scintillator\\
    Detector mass & 20 kton\\
    $\bar{\nu}_e$ Detection Efficiency & 73\%\\
    Running time & 6 years\\
    Thermal power & 36 GW\\
    Energy resolution & 3\% /$\sqrt{\text{E (MeV)}}$\\
    Energy window & 1.8-9 MeV\\
    Number of bins & 200\\
   \end{tabular}
   \caption{\label{tab:juno}JUNO simulated specifications}
   \end{ruledtabular}
\end{table}

Besides T2K-II, \nova-II, and JUNO, we implement a R-SBL neutrino experiment to constrain $\sin^2\theta_{13}$ at $3\%$ uncertainty, which is reachable as prospected in Ref.~\cite{Cao:2016vwh}. This constraint is important to break the parameter degeneracy between \dcp-$\theta_{13}$, which is inherent from the measurement with the electron (anti-)neutrino appearance samples in the A-LBL experiments.\\ 
To calculate the sensitivity, a joint $\chi^2$ is formulated by summing over all individual experiments under consideration without taking any systematic correlation among experiments. For T2K-II and \nova-II, we use a built-in $\chi^2$ function from GLoBES for taking the signal and background normalization systematics with the spectral distortion into account. For JUNO, a Gaussian formula for $\chi^2$ is implemented thanks to a high statistics sample in JUNO. For a given \emph{true} value of the oscillation parameters, $\Vec{\Uptheta}_{\text{truth}}=(\theta_{12},\theta_{13},\theta_{23},\delta_{\text{CP}};\Delta m^{2}_{21}, \Delta m^{2}_{31})_{\text{truth}}$, at a \emph{test} set of oscillation parameters, $\Vec{\Uptheta}_{\text{test}}$, and systematic variations, $\Vec{s}_{\text{syst.}}$, a measure  $\chi^2(\Vec{\Uptheta}_{\text{truth}}| \Vec{\Uptheta}_{\text{test}},\Vec{s}_{\text{syst.}})$ is calculated. It is then minimized over the nuisance parameters (both systematic parameters and marginalized oscillation parameters) to obtain the statistical significance on the hyperplane of parameters of interest. 
\section{\label{sec:results}Results}
Throughout this work, unless otherwise mentioned, we consider the \textit{true} mass hierarchy to be \emph{normal} and oscillation parameters to be as given in Table~\ref{tab:nuoscpara}. Dependence of the MH resolving on the $\theta_{13}$ mixing angle is compensated for in Appendix~\ref{appsec:th13}. In Appendix~\ref{appsec:pot}, as a message to emphasize the vitality of statistics in neutrino experiments, we provide a study on how the total of the T2K-II POT exposure can have a significant impact on the sensitivity results.
\subsection{Determining the neutrino mass hierarchy}
\begin{figure}
\includegraphics[width=0.45\textwidth]{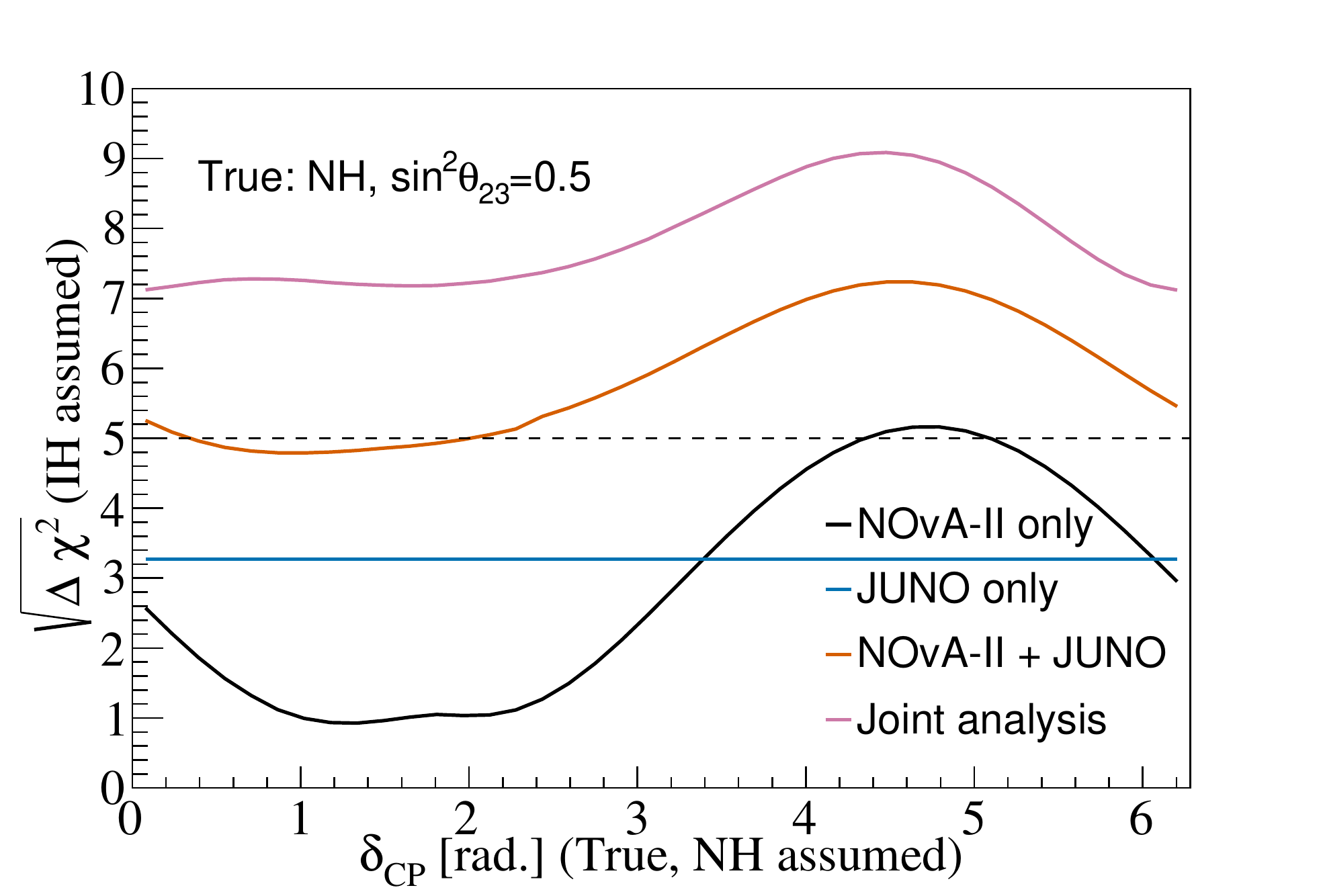}
\includegraphics[width=0.45\textwidth]{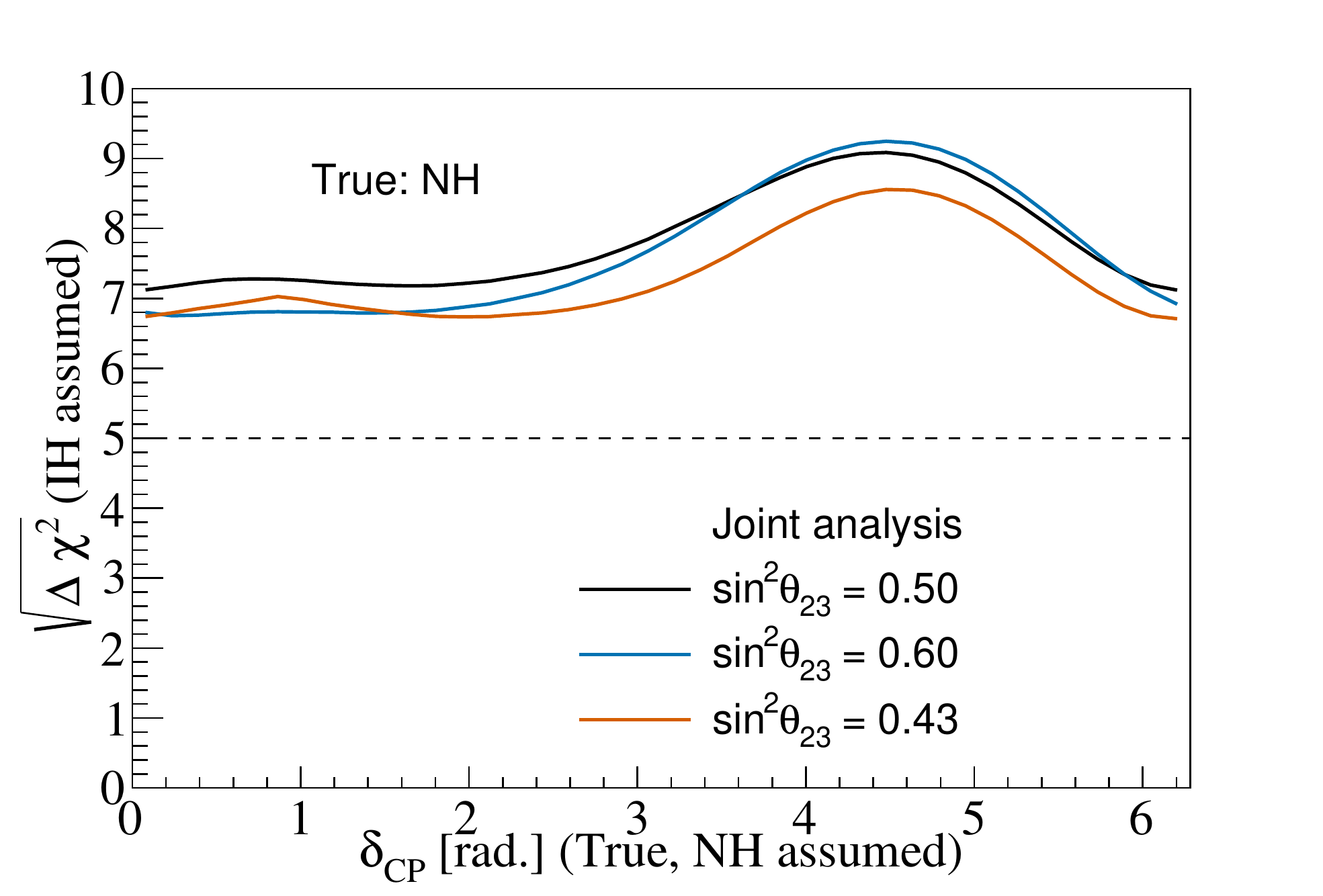}
\includegraphics[width=0.45\textwidth]{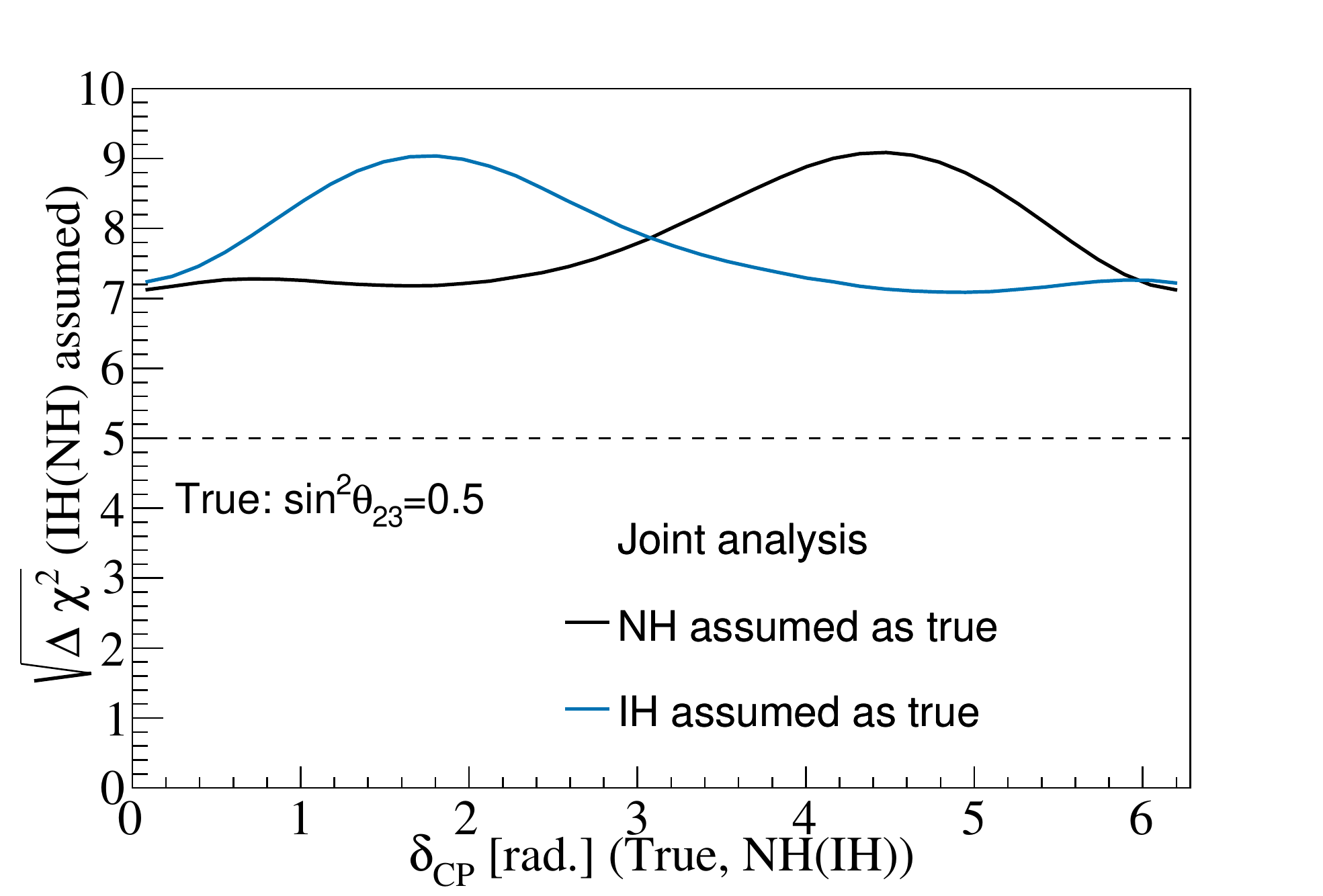}

\caption{\label{fig:sensidmh1d}MH sensitivities as a function of \emph{true} \dcp\ calculated for various experimental setups (top plot); for all considered experiments but at different $\sin^2\theta_{23}$ values (center plot); and for comparing two possible MH hypotheses (bottom plot). For the top and bottom plots, $\sin^{2}\theta_{23}=0.5$ is assumed to be true.}
\end{figure}

To estimate quantitatively the sensitivity of the experiment(s) to the MH determination, we calculate the statistical significance $\sqrt{\Delta \chi^{2}}$ to exclude the \emph{inverted} MH given that the null hypothesis is a \emph{normal} MH, which is indicated by the recent neutrino experiment results. The sensitivity is calculated as a function of \emph{true} \dcp\ since for the A-LBL experiments, the capability to determine the MH depends on the values of the \textit{CP}-violating phase. Technically, for each \emph{true} value of \dcp\ with \emph{normal} MH assumed, marginalized $\chi^{2}$ is calculated for each \emph{test} value of \dcp\ with the MH fixed \emph{to inverted}. Then for each \emph{true} value of \dcp, the minimum value of $\chi^{2}$, which is also equivalent to $\Delta \chi^{2}$ since the \emph{test} value with \emph{normal} MH assumed would give a minimum $\chi^{2}$ close to zero, is obtained. The results, in which we assume $\sin^{2}\theta_{23}=0.5$, are shown in the top plot of Fig.~\ref{fig:sensidmh1d} for different experimental setups: (i) JUNO only; (ii) \nova-II only; (iii) a joint of JUNO and \nova-II; and (iv) a joint of JUNO, \nova-II, T2K-II, and the R-SBL experiment. It is expected that the MH sensitivity of JUNO is more than a 3$\sigma$ \conlev\ and does not depend on \dcp. On the other hand, the \nova-II sensitivity to the MH depends strongly on the \emph{true} value of \dcp. A joint analysis of JUNO with the A-LBL experiments, \nova-II and T2K-II, shows a great boost in the MH determination. This is expected since a joint analysis will break the parameter degeneracy between \dcp\ and the sign of \dmatm. Due to the parameter degeneracy among \dcp\ and the sign of \dmatm, $\theta_{13}$, and $\theta_{23}$ in the measurement with the A-LBL experiments, we also expect that the MH determination depends on the value of $\theta_{23}$. The combined sensitivity of all considered experiments at different values of $\theta_{23}$, (i) maximal mixing at $45^\circ$ ($\sin^{2}\theta_{23}=0.50$), (ii) LO at $41^\circ$ ($\sin^{2}\theta_{23}=0.43$), and (iii) HO at $51^\circ$ ($\sin^{2}\theta_{23}=0.60$), is shown in the center plot of Fig.~\ref{fig:sensidmh1d}. In the bottom plot of Fig.~\ref{fig:sensidmh1d}, we compare the MH sensitivity for two hypotheses: the MH is \emph{normal} and the MH is \emph{inverted}. The result reflects what we expect: (i) the MH resolving with JUNO is less sensitive to its truth since the dominating factor is the separation power between two oscillation frequencies driven by $|\Delta m^2_{31}|$ and $|\Delta m^2_{32}|$, shown in Eq.~(\ref{eq:nuebdis}), and the relatively large mixing angle $\theta_{12}$; and (ii) for the A-LBL experiments like T2K and \nova, the MH is determined through the MH-\dcp\ degeneracy resolving as concisely described in Eq.~(\ref{eq:adcp}). The $A_{\text{CP}}$ amplitude is almost unchanged when one switches from a \emph{normal} MH to an \emph{inverted} MH and simultaneously flips the sign of \dcp. Those results conclude that the \emph{wrong} mass hierarchy can be excluded at a C.L. greater than $5\sigma$ for all the \emph{true} values of \dcp\ and for any value of \thetamu\ in the range constrained by the experiments. In other words, the MH can be determined conclusively by a joint analysis of JUNO with the A-LBL experiments, \nova-II and T2K-II.  
\noindent We find out that in Ref.~\cite{Cabrera:2020own} the authors address a similar objective and come to a quite similar conclusion  even though a different calculation method and assumption of the experimental setup are used. 
\subsection{Unravelling leptonic \textit{CP} violation}
\begin{figure}
\includegraphics[width=0.45\textwidth]{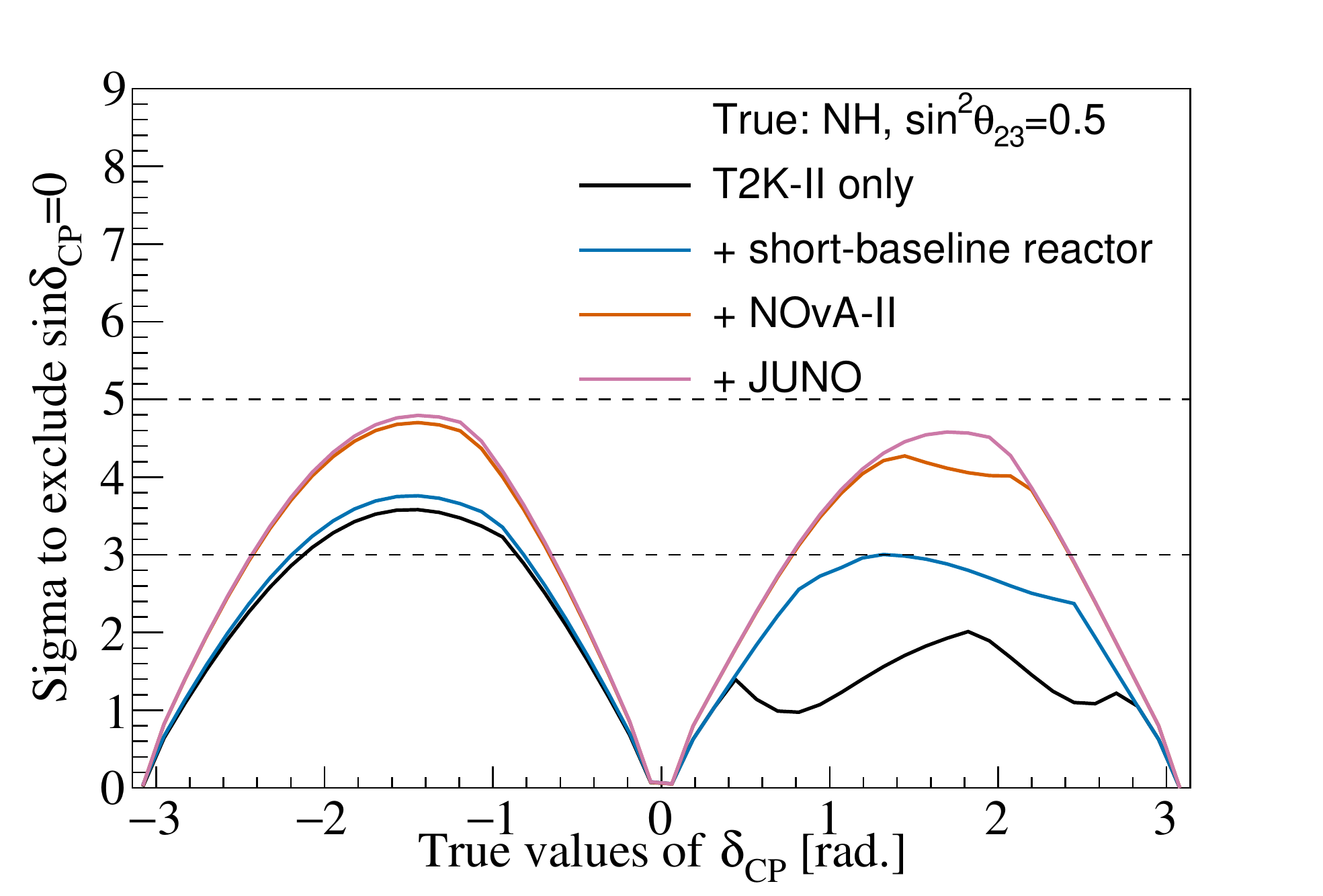}
\includegraphics[width=0.45\textwidth]{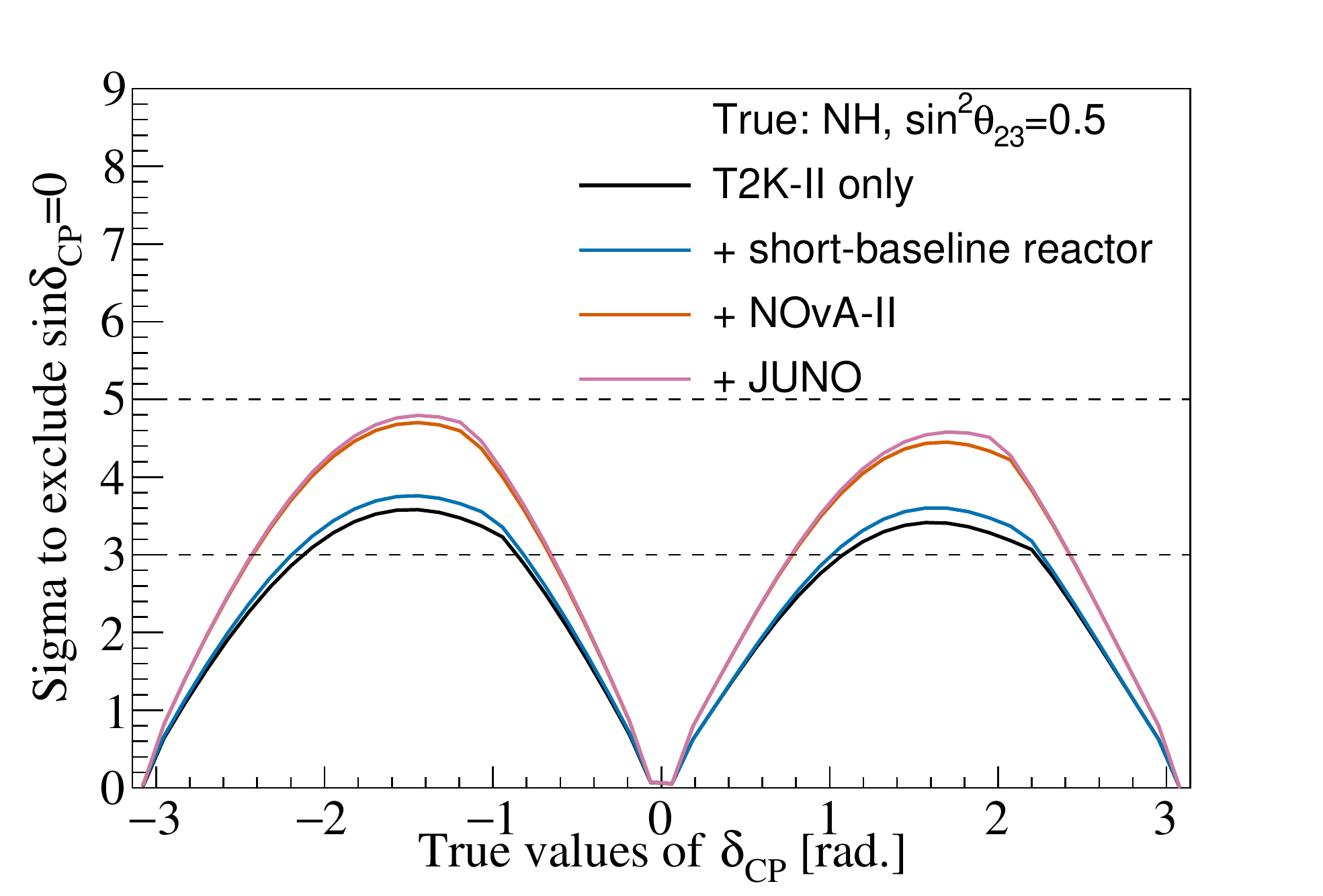}
\caption{\label{fig:sensidcp}CPV sensitivity as a function of the \emph{true} value of \dcp\ obtained with different analyses. \emph{Normal} MH and $\sin^{2}\theta_{23}=0.5$ are assumed to be \emph{true}. The top (bottom) plot is with the MH assumed to be \emph{unknown} (\emph{known}) in the analysis, respectively.}
\end{figure}

\begin{figure}
\includegraphics[width=0.45\textwidth]{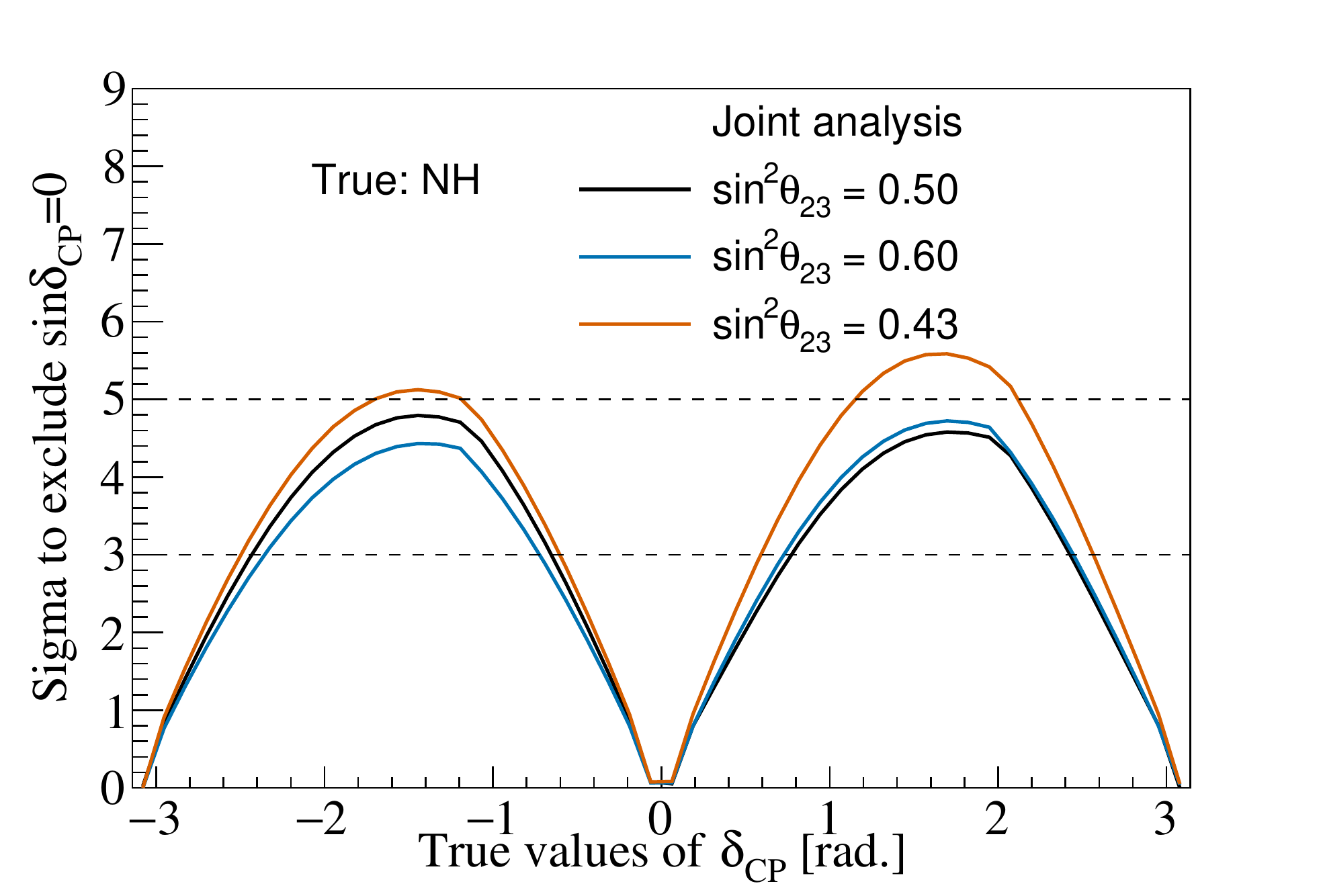}
\includegraphics[width=0.45\textwidth]{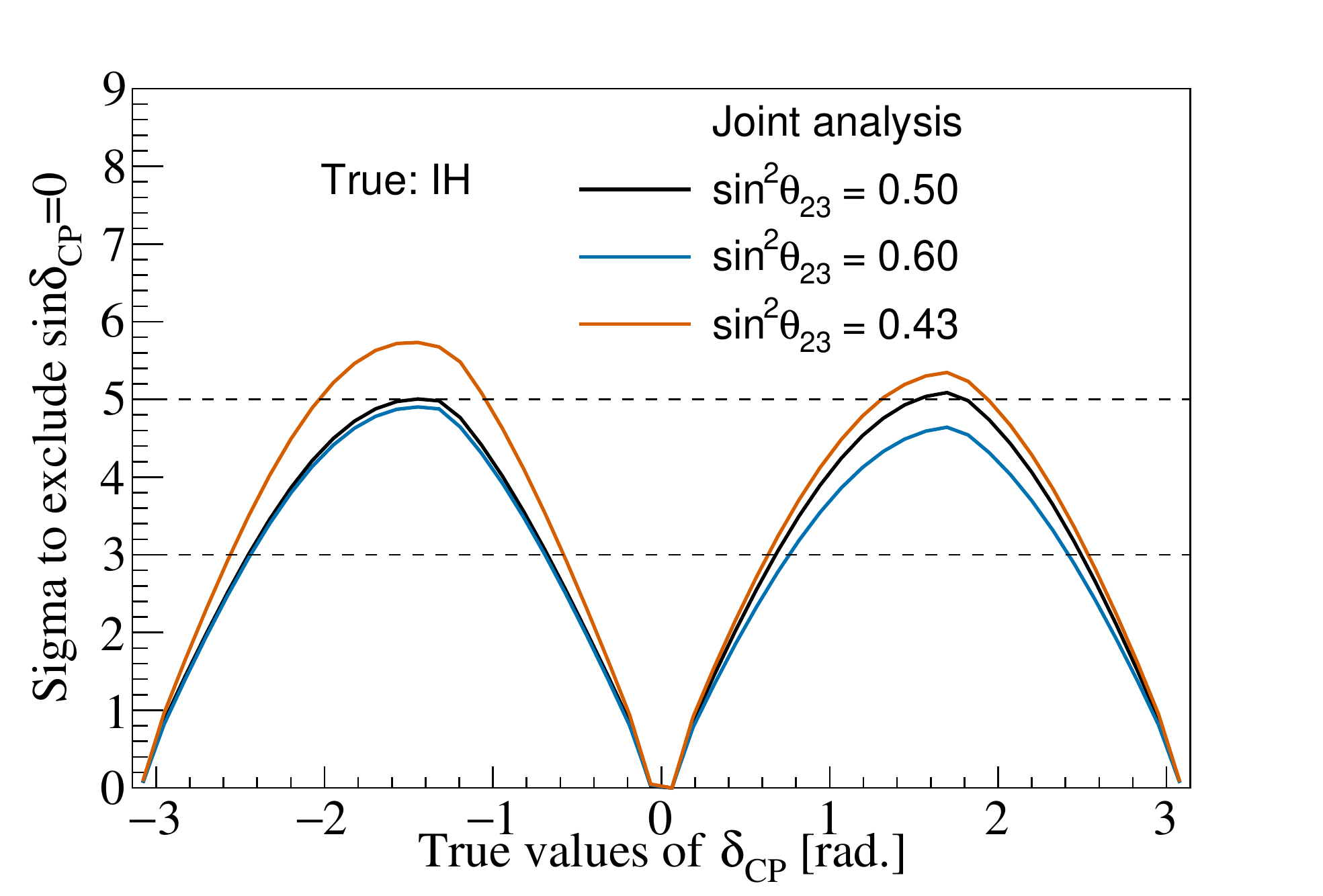}
\caption{\label{fig:sensidcpdifftheta}CPV sensitivity as a function of the \emph{true} value of \dcp\ obtained with a joint analysis of all considered experiments at different \emph{true} $\sin^{2}\theta_{23}$ values (0.43, 0.5, 0.6). The top (bottom) plot is with the \emph{normal} (\emph{inverted}) MH, respectively, and assumed to be true. The MH is assumed to be \emph{unknown} in the analysis.}
\end{figure}

 The statistical significance of $\sqrt{\Delta \chi^2}$ excluding the \textit{CP}-conserving values (\dcp=0,$\pi$) or the sensitivity to CPV is evaluated for any \emph{true} value of \dcp\ with the \emph{normal} MH assumed. For the minimization of $\chi^{2}$ over the MH options, we consider two cases: (i) MH is \emph{known} and \emph{normal}, the same as the truth value, or (ii) MH is \emph{unknown}. Figure~\ref{fig:sensidcp} shows the CPV sensitivity as a function of the \emph{true} value of \dcp\ for both MH options obtained by different analyses: (i) T2K-II only; (ii) a joint of T2K-II and R-SBL experiments; (iii) a joint of T2K-II, \nova-II, and R-SBL experiments; and (iv) a joint of T2K-II, \nova-II, JUNO, and R-SBL experiments. The result shows that whether the MH is \emph{known} or \emph{unknown} affects the first three analyses, but not the fourth. This is because, as concluded in the above section, the MH can be determined conclusively with a joint analysis of all considered experiments. 
\noindent It can be seen that the sensitivity to \textit{CP} violation is driven by T2K-II and \nova-II. Contribution of the R-SBL neutrino experiment is significant only at the region where \dcp\ is between 0 and $\pi$ and when the MH is not determined conclusively. JUNO further enhances the CPV sensitivity by lifting up the overall MH sensitivity and consequently breaking the MH-\dcp\  degeneracy. At \dcp\ close to $-\pi/2$, which is indicated by recent T2K data \cite{Abe:2019vii}, the sensitivity of the joint analysis with all considered experiments can approximately reach a 5$\sigma$ \conlev\ We also calculate the statistical significance of the CPV sensitivity as a function of \emph{true} \dcp\ at different values of \thetamu, as shown in Fig.~\ref{fig:sensidcpdifftheta}. When an \emph{inverted} MH is assumed, although $A_{\text{CP}}$ amplitude  fluctuates in the same range as with a \emph{normal} MH, the probability and rate of $\nu_e$ appearance becomes smaller to make the statistic error, $\sigma_{\nu_e}^{\text{stat.}}$, lower. In sum, sensitivity to \textit{CP} violation, which is proportional to $A_{\text{CP}}/\sigma_{\nu_e}^{\text{stat.}}$,  is slightly higher if the \emph{inverted} MH is assumed to be \emph{true} as shown at the bottom of the Fig.~\ref{fig:sensidcpdifftheta}.
Table~\ref{tab:cpth23} shows the fractional region of all possible \emph{true} $\delta_{\text{CP}}$ values for which we can exclude \textit{CP}-conserving values of $\delta_{\text{CP}}$ to at least a $3\sigma$~\conlev, obtained by the joint analysis of all considered experiments. Due to the fact that the MH is resolved completely with the joint analysis, the CPV sensitivities are quantitatively identical no matter whether the MH is assumed to be \emph{known} or \emph{unknown}. 
\begin{table}
    \begin{ruledtabular}
    \caption{\label{tab:cpth23}Fractional region of $\delta_{\text{CP}}$, depending on $\sin^{2}\theta_{23}$, can be explored with $3\sigma$ or higher significance}
 \begin{tabular}{l|ccc}
    Value of $\sin^{2}\theta_{23}$ & 0.43 & 0.50 & 0.60\\\hline
    Fraction of \emph{true} $\delta_{\text{CP}}$ values (\%), NH & 61.6 & 54.6 & 53.3\\
    Fraction of \emph{true} $\delta_{\text{CP}}$ values (\%), IH & 61.7 & 57.2 & 54.2\\
   \end{tabular}
  \end{ruledtabular}
\end{table}

\subsection{Precision measurement of other oscillation parameters}
\begin{figure}
\begin{subfigure}{.48\textwidth}
\includegraphics[width=0.95\textwidth]{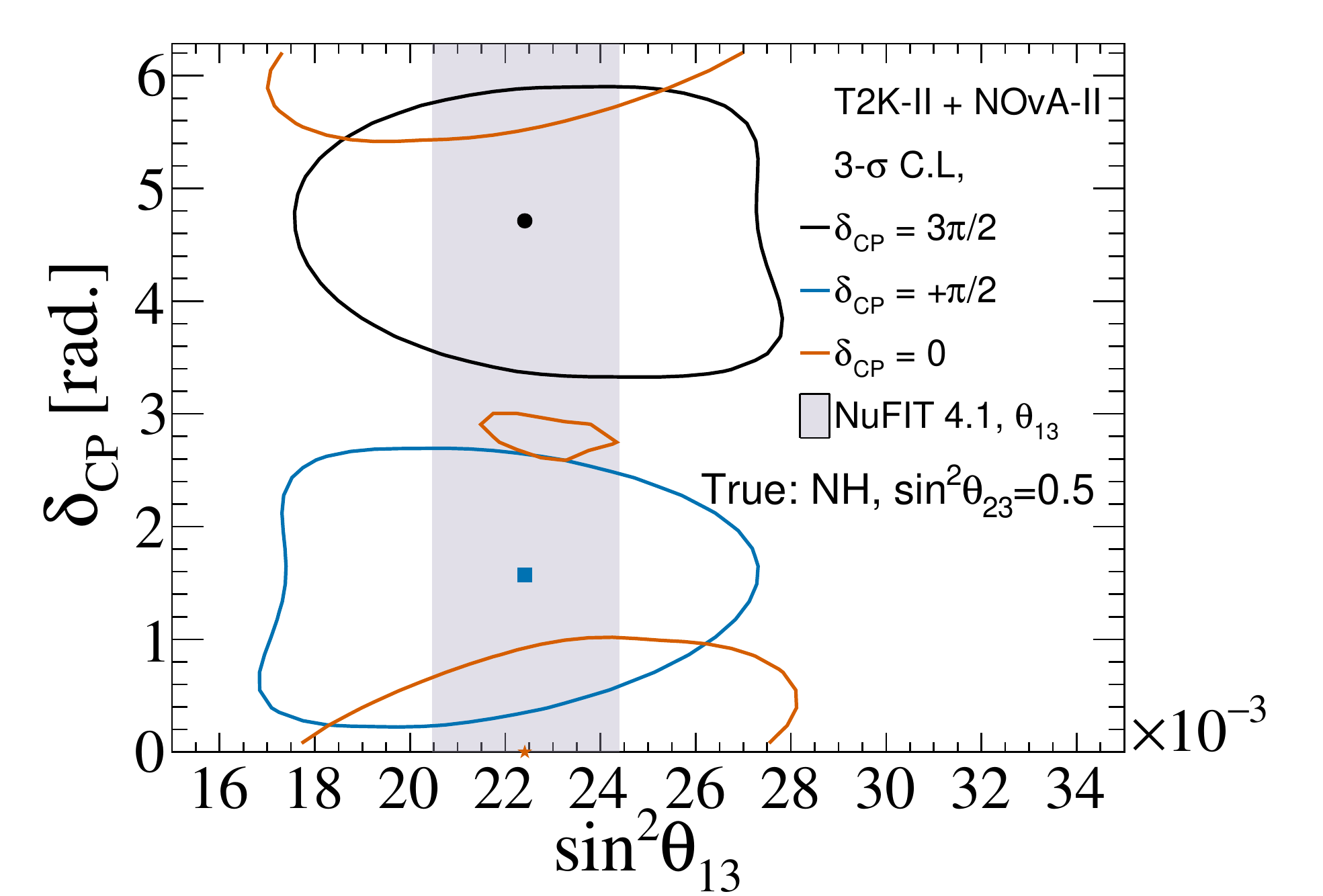} 
  \caption{\label{fig:13lblreac}}
\end{subfigure}
\begin{subfigure}{.48\textwidth}
\includegraphics[width=0.95\textwidth]{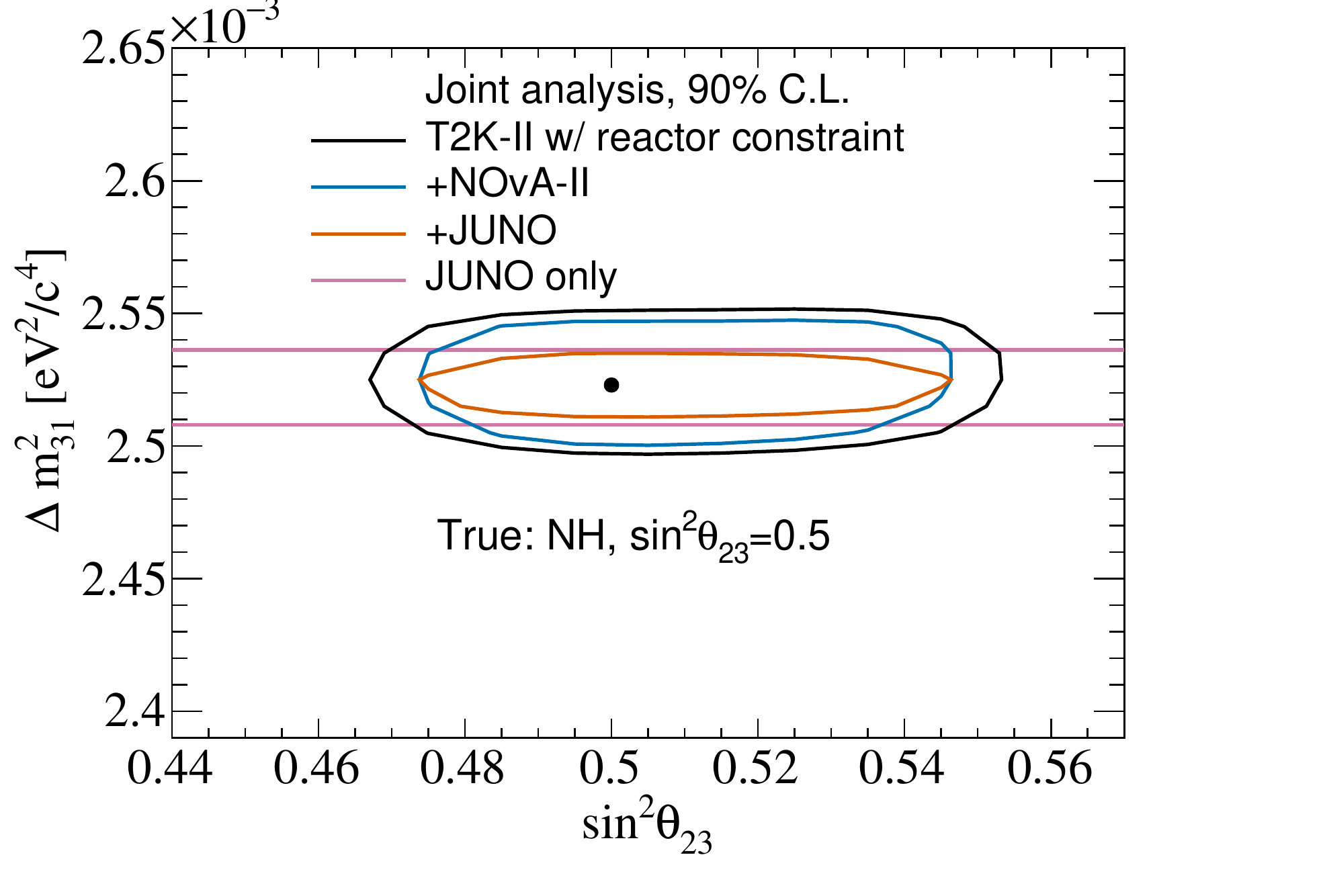} 
\caption{\label{fig:dm31precise}}
\end{subfigure}%
\caption{\label{fig:precision}(a) Allowed region of $\sin^{2}\theta_{13}$-\dcp\ at a 3$\sigma$ \conlev\ compared with a joint analysis of T2K-II and \nova-II and the present constraint from the global data~\cite{esteban2019global}. (b) Allowed region in the $\sin^{2}\theta_{23}-\Delta m^{2}_{31}$ space at a 90\% C.L. with various experimental setups. \emph{Normal} MH and $\sin^{2}\theta_{23}=0.5$ are assumed to be true. }

\end{figure}

\paragraph{\textbf{$\theta_{13}$ mixing angle and atmospheric oscillation parameters:}}
The \textbf{$\theta_{13}$} mixing angle can be constrained precisely by measuring the disappearance of $\overline{\nu}_{e}$ in the  R-SBL neutrino experiment. The A-LBL experiments, on the other hand, can provide a constraint of the \textbf{$\theta_{13}$} mixing angle correlated to \dcp, mainly thanks to the measurements of the appearance of $\nu_{e}(\overline{\nu}_{e})$ from the beam of $\nu_{\mu}(\overline{\nu}_{\mu})$, respectively. The sensitivities are calculated at three different \emph{true} values of \dcp\ $(0,\pm\frac{\pi}{2})$.
A 3$\sigma$~\conlev\ range of $\sin^2\theta_{13}$ $[0.02046,0.02440]$ is taken from Ref.~\cite{esteban2019global}. Figure~\ref{fig:13lblreac} shows a 3$\sigma$ \conlev\ allowed region of $\sin^{2}\theta_{13}$-\dcp\ obtained with a joint analysis of the T2K-II and \nova-II experiments. The precision of $\sin^2\theta_{13}$ can be achieved between 6.5$\%$ and 10.7$\%$ depending on the \emph{true} value of \dcp. It will be interesting to compare the measurements of \textbf{$\theta_{13}$} from R-SBL experiments and from the A-LBL experiments with such high precision.

\begin{figure}
\begin{subfigure}{.48\textwidth}
\centering
\includegraphics[width=0.95\textwidth]{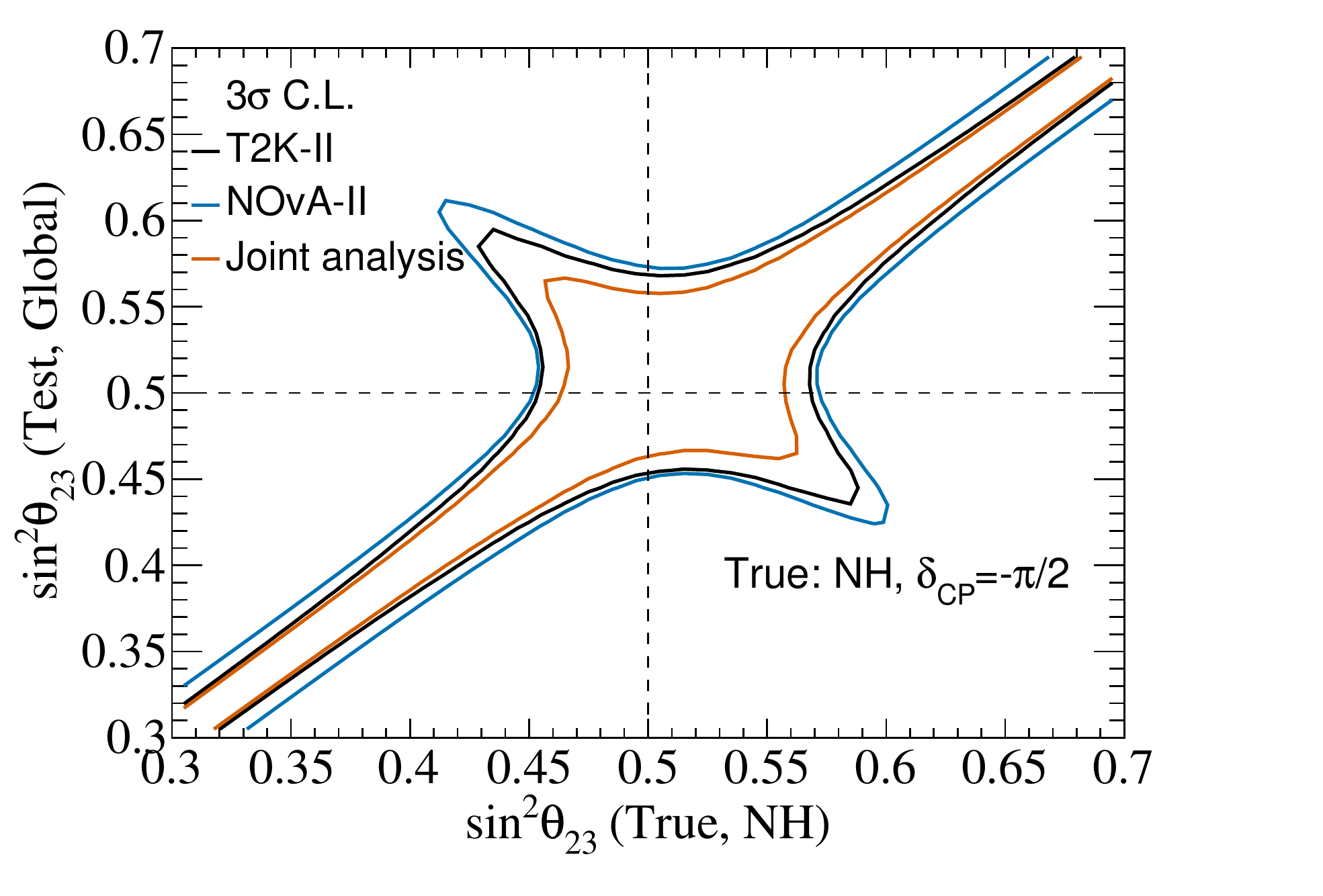}
  \caption{}
  \label{fig:sensith23comb3sigma}
\end{subfigure}
\begin{subfigure}{.48\textwidth}
\centering
\includegraphics[width=0.95\textwidth]{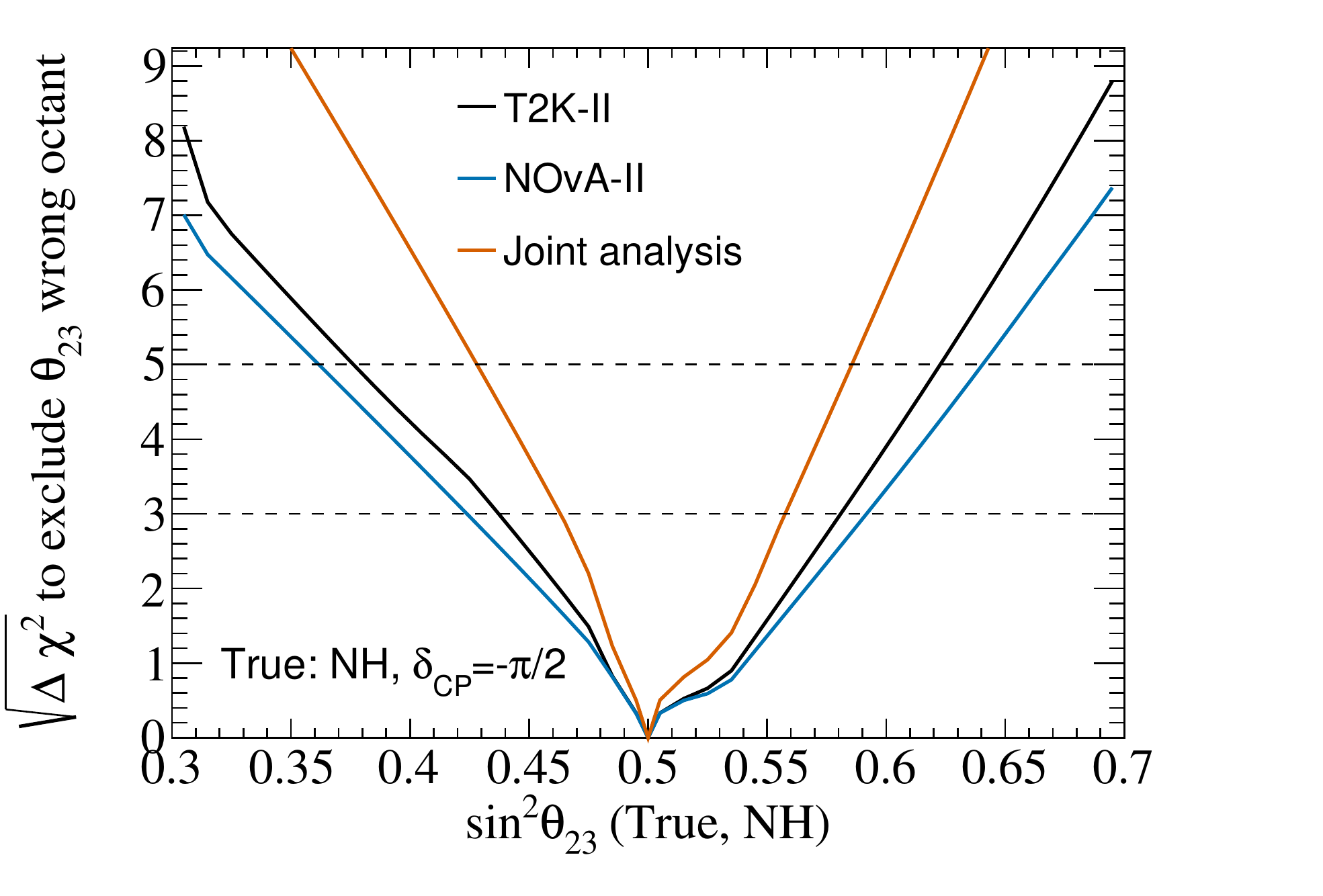} 
\caption{}
  \label{fig:th23octantresolving}
\end{subfigure}%
\caption{Allowed region of $\sin^2\theta_{23}$ at a 3$\sigma$ C.L (a) and the statistical significance of excluding the \emph{wrong} octant as a function of $\sin^2\theta_{23}$ (b). \emph{Normal} MH and $\delta_{\text{CP}} =-\frac{\pi}{2}$ are assumed to be true.}
\label{fig:octantresolving}
\end{figure}

As shown in Fig.~\ref{fig:dm31precise}, both JUNO alone and a combined sensitivity of T2K-II and \nova-II experiments can reach a sub-percent-level precision on the atmospheric mass-squared splitting \dmatm. A comparison at such precision may provide a very good test for the PMNS framework. Besides, assuming a maximal mixing $\sin^{2}\theta_{23}=0.5$, a combined sensitivity of T2K-II and \nova-II can achieve approximately 6$\%$ and 3$\%$ precision for the upper and lower limit on $\sin^{2}\theta_{23}$. A capability to solve the $\theta_{23}$ octant in case the mixing angle $\theta_{23}$ is not maximal is discussed below.
\paragraph{\textbf{Resolving the octant of the $\theta_{23}$ mixing angle:}}
We consider a range [0.3, 0.7] of possible \emph{true} $\sin^2\theta_{23}$ values and that the \emph{true} MH is \emph{normal}. For each \emph{true} $\sin^2\theta_{23}$ value, the marginalized $\chi^2$ is calculated at various values of \emph{test} value $\theta_{23}$ with both possibilities of the MH. The minimization over the MH options is firstly performed to obtain global minimum $\chi^2$ for any combination of the \emph{true} and \emph{test} values of $\theta_{23}$. The allowed regions of $\sin^2\theta_{23}$ as a function of $\sin^2\theta_{23}$ can be obtained, e.g., at a 3$\sigma$ C.L, as shown in Fig.~\ref{fig:sensith23comb3sigma}. The statistical significance to exclude the \emph{wrong} octant given a \emph{true} (nonmaximal) value of $\theta_{23}$ is calculated by taking the difference between the mimimal value of the global $\chi^2$ in the \emph{wrong} octant and the \emph{true} octant of $\theta_{23}$. The octant resolving sensitivities with T2K-II, \nova-II alone, or with a combined analysis is shown in Fig.~\ref{fig:th23octantresolving}. The $\theta_{23}$ octant resolving power can be enhanced significantly when combining T2K-II and \nova-II data samples, particularly the $\theta_{23}$ octant can be determined at a 3$\sigma$ C.L. or higher if $\sin^2\theta_{23}$ is $\leq0.46$ or $\geq0.56$. 

\subsection{Discussion}
We briefly discuss the implications that have arisen from our results in light of the recent updated results from T2K~\cite{patrick_dunne_2020_4154355}, NOvA~\cite{alex_himmel_2020_4142045}, SK~\cite{yasuhiro_nakajima_2020_4134680}, IceCube DeepCore~\cite{summer_blot_2020_4156203}, and MINOS(+)~\cite{thomas_carroll_2020_4156118} presented at the Neutrino 2020 conference. T2K prefers the \emph{normal} MH with a Bayes factor of 3.4; SK disfavors the \emph{inverted} MH at a 71.4\textup{--}90.3$\%$ C.L.; both \nova\ and MINOS(+) disfavor the \emph{inverted} MH at a C.L. less than 1$\sigma$. The prospect of completely resolving the MH by combining T2K-II, \nova-II, and JUNO by 2027 is thus very encouraging. On the leptonic CPV search, the leading measurement is from T2K where $35\%$ of \dcp\ values are excluded at a $3\sigma$ C.L. Comparing this to Ref.~\cite{Abe:2019vii}, although the statistic significance of excluding \textit{CP} conservation is reduced from a 95$\%$ C.L. to a 90$\%$ C.L., the updated data looks more consistent with the PMNS prediction than before. While SK also favors the maximum \textit{CP} violation, \nova\ shows no indication of asymmetry of neutrino and antineutrino behaviors. With the combined analysis of T2K-II, \nova-II, and JUNO by 2027, it is expected that more than half of the \dcp\ values can be excluded with more than a $3\sigma$ C.L. If the \emph{true} \dcp\ is near $\delta_{\text{CP}}=\pm\frac{\pi}{2}$, discovery of the leptonic CPV with a $5\sigma$ C.L. is within reach. Regarding the octant of the \thetamu\ mixing angles, T2K, \nova, SK, and MINOS(+) data prefer nonmaximum with statistic significance between a 0.5$\sigma$ to 1.5$\sigma$ C.L. If the \emph{true} value of \thetamu\ is close to the best fit in the global data fit~\cite{Esteban:2020cvm}, \thetamu$=0.57$, a combined analysis of T2K-II, \nova-II, and JUNO can exclude the \emph{wrong} octant with a 3$\sigma$ C.L. There is a room for improvement in the above-mentioned physic potentials, for example, by adding an atmospheric neutrino data sample from the SK experiment. There are ongoing efforts to combine data from T2K and SK along with a joint analysis of T2K and \nova. Such activities are vital to realizing a grand framework for combining the special-but-statistically-limited neutrino data in the future.

\section{\label{sec:fin}Conclusion}
We have studied the physics potentials of a combined analysis of the two accelerator-based long-baseline experiments, T2K-II and \nova-II, and a reactor-based medium-baseline experiment, JUNO. We have shown that the combined analysis will unambiguously determine the neutrino mass hierarchy given any \emph{true} values of \dcp\ and \thetamu\ within the present allowed range. The combined analysis provides a very appealing sensitivity for the leptonic \textit{CP} violation search. Particularly, \textit{CP}-conserving values of \dcp\ can be excluded with at least a 3$\sigma$ \conlev\ for 50\% of all the possible \emph{true} values of \dcp. At \textit{CP} violation phase values close to $\delta_{\text{CP}}=\pm\frac{\pi}{2}$, a discovery of \textit{CP} violation in the leptonic sector at the $\sim 5\sigma$ \conlev\ becomes possible. Besides, a combined analysis of T2K-II and \nova-II can reach a few percent precision on the $\theta_{13}$ mixing angle and sub-percent-level precision on the \dmatm mass-squared splitting, which can provide interesting tests of the standard PMNS framework by comparing the results to measurements from reactor-based short-baseline neutrino experiments and JUNO, respectively. Also, a combined analysis of T2K-II and \nova-II offers a great sensitivity to determine the octant of the \thetamu\ mixing angle.

Finally, we would like to emphasize that the joint analysis in reality is foreseen to be more complicated than what we have done. Many systematic sources must be taken into account for each experiment and for a joint analysis; the correlation of systematic errors among experiments are important for extracting precisely the oscillation parameters. However, we affirm that the above conclusions are still valid since the measurement uncertainties, particularly for \textit{CP} violation and the neutrino mass hierarchy, are still dominated by statistical errors.
\section*{Acknowledgements}
The authors thank the selection committee of the XXIX International Conference on Neutrino Physics and Astrophysics 2020, Fermilab for giving an opportunity to present the preliminary results of this work. A.N. thanks the International Centre for Interdisciplinary Science and Education (ICISE), Quy Nhon, Vietnam for the warm hospitality to carry out the initial part of the work. The research of T.V.Ngoc and N.T.H.Van is funded by the National Foundation for Science and Technology Development (NAFOSTED) of Vietnam under Grant No. 103.99-2020.50. N.K.F and A.N. acknowledge DST-SERB, Government of India, for the research project vide Grant No. EMR/2015/001683 (for the work).

\appendix

\section{Dependence of mass hierarchy determination on $\theta_{13}$}\label{appsec:th13}
\begin{figure}
\includegraphics[width=0.9\linewidth]{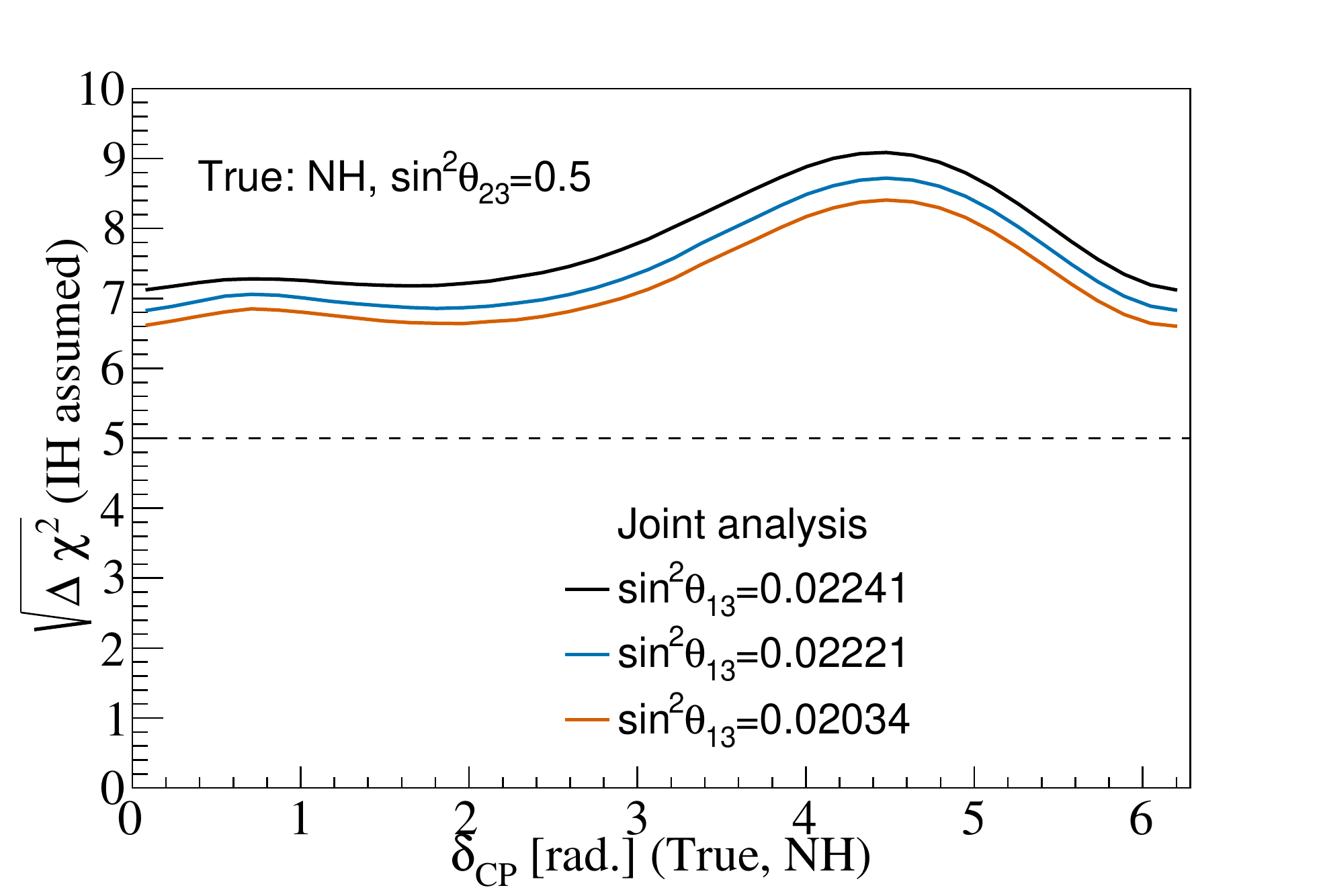}
\caption{\label{fig:diffth13}Dependence of the neutrino MH sensitivity on the $\theta_{13}$ true values: $\sin^2\theta_{13} = 0.02241$ is the best fit obtained with NuFIT 4.1~\cite{esteban2019global}, $\sin^2\theta_{13} = 0.02221$ is with NuFIT 5.0~\cite{Esteban:2020cvm}, and $\sin^2\theta_{13} = 0.02034$ is a 3$\sigma$ C.L. lower limit. \emph{Normal} MH and $\sin^{2}\theta_{23}=0.5$ are assumed to be true. }

\end{figure}

As pointed out in Ref.~\cite{Suekane:2015yta}, the CPV sensitivity with the A-LBL neutrino experiments does not depend on the the true value of $\theta_{13}$. However this is not the case for the MH sensitivity since the $\overline{\nu}_e$ disappearance rate in JUNO is proportional to $\sin^22\theta_{13}$ as shown in Eq.~(\ref{eq:nuebdis}). This feature is presented in Fig.~\ref{fig:diffth13} where the sensitivities of the neutrino MH are studied with three different values of $\sin^2\theta_{13}$: $\sin^2\theta_{13} = 0.02241$ is the best fit obtained with NuFIT 4.1~\cite{esteban2019global}, $\sin^2\theta_{13} = 0.02221$ is with NuFIT 5.0~\cite{Esteban:2020cvm}, and $\sin^2\theta_{13} = 0.02034$ is a 3$\sigma$ C.L. lower limit. Although the neutrino MH sensitivity is slightly reduced with smaller values of $\sin^2\theta_{13}$, the MH resolution is still well above a 5$\sigma$ C.L. 

\section{Sensitivity with different scenarios of the T2K-II POT exposure}\label{appsec:pot}
\begin{figure}
\begin{subfigure}{.48\textwidth}
\includegraphics[width=0.95\textwidth]{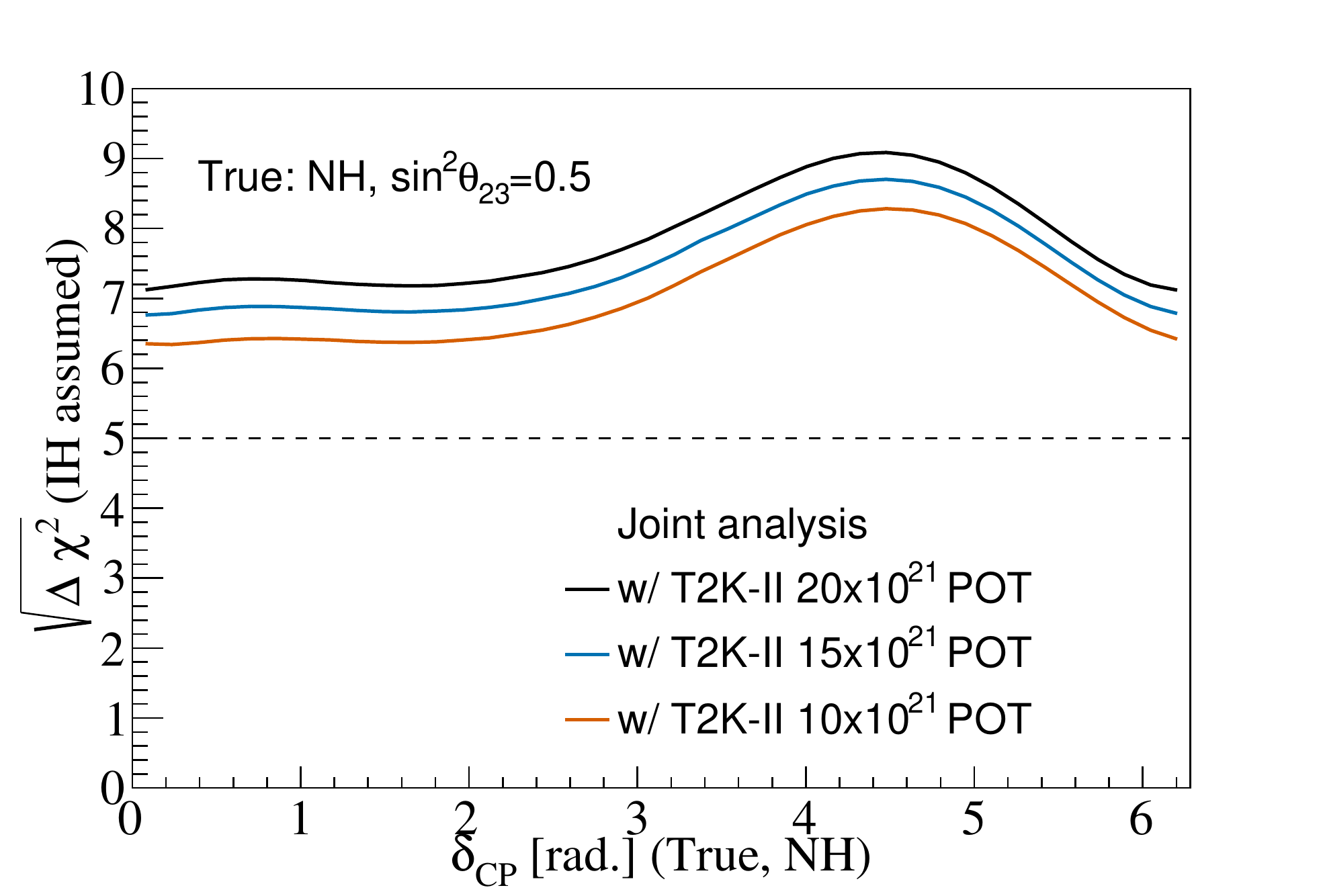}
  \caption{\label{fig:t2kpotmh}}
\end{subfigure}
\begin{subfigure}{.48\textwidth}
\includegraphics[width=0.95\textwidth]{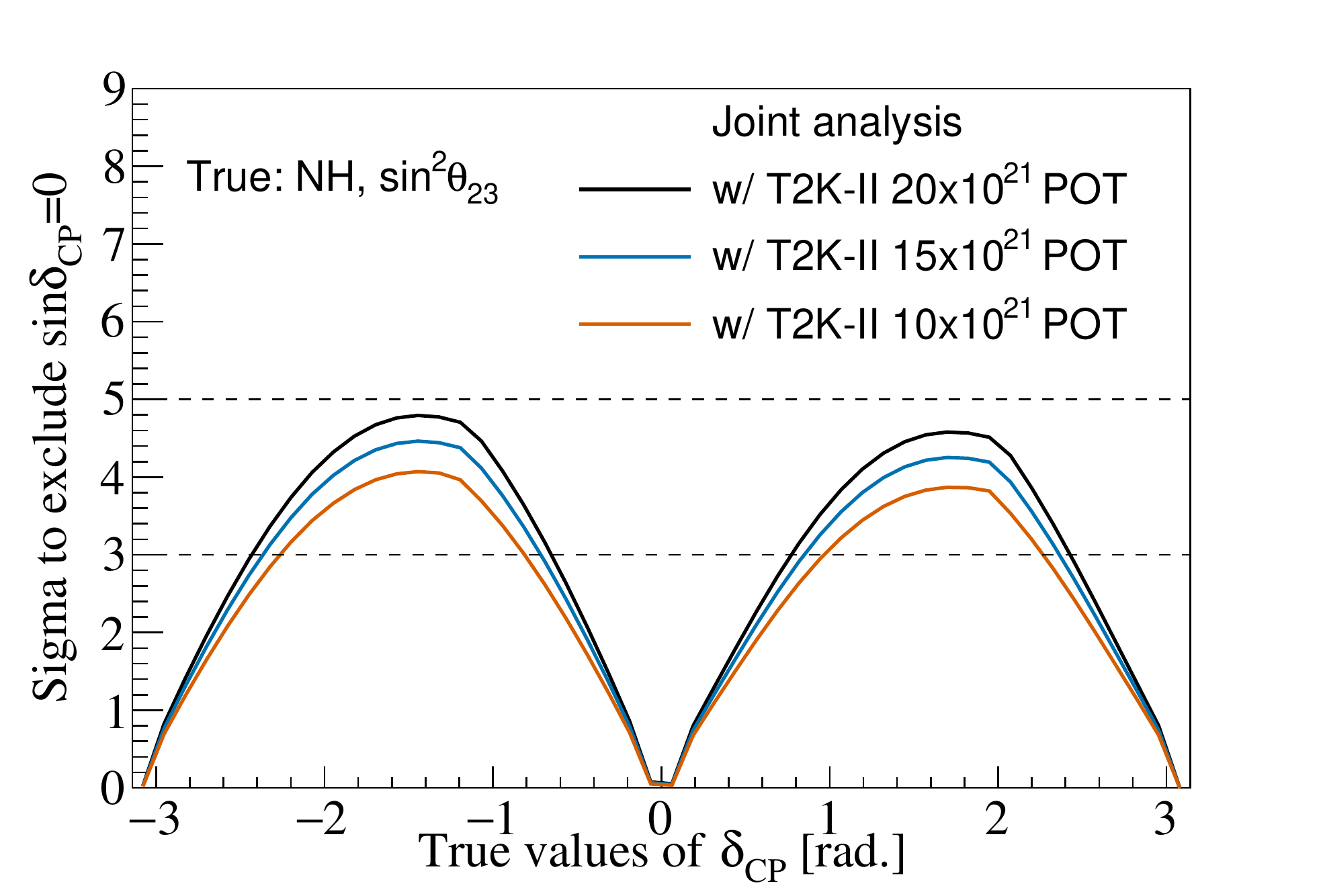}
\caption{\label{fig:t2kpotdcp}}
\end{subfigure}%
\caption{\label{fig:t2kpot} Dependence of the combined sensitivity on T2K-II POT exposure. (a) MH sensitivities as a function of \emph{true} \dcp. (b) CPV sensitivity as a function of the \emph{true} value of \dcp\ obtained with a joint analysis of all considered experiments. \emph{Normal} MH and $\sin^{2}\theta_{23}=0.5$ are assumed to be true.}

\end{figure}

Due to the budget issue, it is possible that T2K-II will take data less than the original proposal as discussed in Ref.~\cite{Cabrera:2020own}. In this sense, we study three scenarios of the T2K-II POT exposure: $20\times 10^{21}$, $15\times 10^{21}$, and $10\times 10^{21}$ POT. While the MH resolving is still well above a 5$\sigma$ C.L., the CPV sensitivity depends significantly on the POT exposure as shown in Fig.~\ref{fig:t2kpot}. However there is still a large fraction of \dcp\ value excluded with a 3$\sigma$ C.L. The study emphasizes the importance of providing as many proton beams as possible to the T2K experiment for reaching the highest capability of CPV search.

\bibliography{apssamp}

\end{document}